\documentclass[onecolumn,aps,prd,preprintnumbers]{revtex4}

\usepackage{graphicx}
\usepackage{dcolumn}
\usepackage{bm}
\usepackage{subfigure}
\usepackage{slashed}
\usepackage{marvosym}
\usepackage{textcomp}
\usepackage{threeparttable}
\usepackage{booktabs}
\usepackage{setspace}
\usepackage{latexsym}
\usepackage{mciteplus}
\renewcommand{\TPTtagStyle}%
{\normalsize\textit}

\begin{document}

\title{Transverse flow induced by inhomogeneous magnetic fields in the Bjorken expansion}
\author{Shi Pu$^1$, Di-Lun Yang$^2$\footnote{dy29@phy.duke.edu}}
\affiliation{$^1$Institute for Theoretical Physics, Goethe University,
Max-von-Laue-Str. 1, 60438 Frankfurt am Main, Germany.\\
$^2$Theoretical Research Division, Nishina Center, RIKEN, Wako, Saitama 351-0198, Japan.}
\date{\today}
\begin{abstract}
We investigate the magnetohydrodynamics in the presence of an external magnetic field 
following the power-law decay in proper time and having spatial inhomogeneity characterized by a 
Gaussian distribution in one of transverse coordinates under the Bjorken expansion. 
The leading-order solution is obtained in the weak-field approximation, where both energy 
density and fluid velocity are modified. It is found that the spatial gradient of the magnetic 
field results in transverse flow, where the flow direction depends on the decay exponents 
of the magnetic field. We suggest that such a magnetic-field-induced effect might influence anisotropic flow in heavy ion collisions. 
\end{abstract}
\maketitle
\section{Introduction \label{introduction}}
Recently, the influence of strong magnetic and electric fields on the hot and dense matter, such as the quark-gluon plasma (QGP) created in relativistic nucleus-nucleus collisions, has been intensively studied. The fast-moving nuclei in peripheral collisions could produce extremely strong magnetic fields of the order of $B\sim10^{18}-10^{19}G$ in early times \cite{Gyulassy:2004zy,Bzdak:2011yy,Deng:2012pc}. 
It has been proposed that such strong magnetic fields could affect the thermal-photon emission \cite{Tuchin:2010gx,Basar:2012bp,Wu:2013qja,Muller:2013ila} and heavy flavor physics including heavy quarkonium production \cite{Yang:2011cz,Machado:2013rta,Alford2013a,Guo:2015nsa} and heavy-quark diffusion \cite{Fukushima:2015wck} in the QGP. Moreover, in the presence of chirality imbalance, the magnetic fields may induce charge currents and density waves, namely
the chiral magnetic effect (CME) \cite{Kharzeev:2007jp,Fukushima:2008xe}
and chiral magnetic waves (CMW) \cite{Kharzeev:2010gd}. The effects
are one of candidates to understand the asymmetry in the angular distribution
of charge particles and difference of the elliptic flows of $\pi^{\pm}$
\cite{Burnier:2011bf}. Those phenomena can be interpreted in the
language of Berry phase and the effective chiral kinetic equations
(CKE), which are obtained by path integral \cite{Stephanov:2012ki,Chen:2013iga,Chen:2014cla},
Hamiltonian approaches \cite{Son:2012wh,Son:2012zy} and quantum kinetic
theory via Wigner functions \cite{Gao:2012ix,Chen:2012ca}. A series
of reviews and more references can be found in Ref. \cite{Bzdak:2012ia,Kharzeev:2013ffa,Kharzeev:2015kna}. In addition to strong magnetic fields, chiral currents and density waves
can also be induced by electric fields, named the chiral electric separation
effect (CESE) \cite{Huang:2013iia,Pu:2014cwa,Jiang:2014ura,Pu:2014fva}.
Furthermore, in the presence of electric fields perpendicular to the
magnetic fields, chiral Hall currents are also expected, called the chiral
Hall separation effect (CHSE) \cite{Pu:2014fva}, which might cause
the asymmetric charge and chirality distribution in rapidity. Those
phenomena have drawn lots of attention to the studies of hot and dense
matter under the influence of strong magnetic fields. 

As a very popular and triumphal tool, relativistic hydrodynamics
has been widely used to study heavy ion collisions (e.g. see Ref. \cite{Romatschke2007,Luzum2008,Song2008b,Song:2008si,Schenke:2011bn,Roy:2012jb,Niemi:2012ry}).
In order to investigate those charge and chiral separation effects,
one will consider the combination of relativistic hydrodynamic equation
and Maxwell's equations, i.e. relativistic magneto-hydrodynamics (MHD).
Although the strong magnetic fields decay rapidly in the vacuum \cite{Kharzeev:2007jp}
and substantially delayed in the presence of a electrically conducting
media \cite{Tuchin:2013apa,Gursoy:2014aka,Zakharov:2014dia, Li:2016tel}, recent
studies from event-by-event simulations show the magnetic
energy density can be comparable to fluid energy density  in some
events at $\sqrt{s_{NN}}=200\textrm{ GeV}$ with the impact parameters
$b\sim10$fm \cite{Roy:2015coa}. Therefore, it is still worthy to study relativistic
MHD in relativistic heavy ion collisions. To this scope, a numerical
code of solving 3+1 dimensional MHD is required. 

Notwithstanding numerical simulations of hydrodynamics successfully describe numbers of 
experimental measurements, many analytic studies which aim at mimicking nucleus-nucleus 
collisions have been attempted in order to acquire deeper understandings for numerical results. 
Following two renown solutions with one-dimensional expansion found by Landau \cite{Landau:1953gs} and Bjorken \cite{Bjorken:1982qr}, one of recent improvements is made by Gubser to incorporate the transverse expansion on top of the Bjorken's solution for conformal fluids \cite{Gubser:2010ze,Gubser:2010ui}. Based on the approach in Gubser's solution, refined solutions have been found \cite{Hatta:2014gga,Hatta:2014gqa} and applied to evaluate anisotropic flow in comparison with experimental data \cite{Hatta:2014jva,Pang:2014ipa}. Along the same direction, an analytic solution in 3+1 dimensional hydrodynamics with rapidity dependence has been introduced recently \cite{Hatta:2015ldk}. In the same spirit, we would like to investigate some well-known hydrodynamic models in the framework of MHD and seek for analytic solutions. The study was initiated by one of the authors of this paper, where an one-dimensional fluid following
longitudinal boost-invariant expansion as the Bjorken flow
with a transverse and time-dependent magnetic field has been investigated \cite{Roy:2015kma}. In ideal MHD limits,
i.e. the infinite conductivity and no dissipative effects, it is remarkable
that the decay of energy density is the same as the case without magnetic
fields because of ``frozen-flux theorem'' \cite{rezzolla2013,Landau}.
In Ref.\cite{Pu:2016ayh}, the magnetization effect is added to the Bjorken
flow with MHD. Also see Ref.\cite{Pang:2016yuh}, where the authors considered 3+1 D numerical hydrodynamics with an effective source driven by magnetization. In the presence of an external homogeneous 
magnetic field in a power-law decay $\sim \tau^{-a}$ with $\tau$ being proper time and $a$ being an arbitrary number, 
the solutions are distinguished between the scenarios in which the magnetic
field decays more slowly or more rapidly than in the ideal-MHD case, where the former corresponds to $a < 1$ and the latter corresponds to $a > 1$. For the case $a=1$, it goes back to the ideal MHD. 
In the first scenario, the decay of energy density
is faster than the case without mangetic field. While, in the second scenario, the decay of energy 
density slows down.

In this work, we will consider the system with an inhomogeneous external magnetic
field compared to the previous case with a homogeneous one in the
transverse plane \cite{Roy:2015kma}. Since the energy density of the fluid is modified as shown in the
 homogeneous case \cite{Roy:2015kma}, one may intuitively expect that the spatial gradient of the
 magnetic field may further induce inhomogeneity of the energy distribution and engender anisotropic
 flow in the transverse plane. For the sake of simplicity, we assume the external magnetic fields
are small compared to the fluid energy density.
Therefore, we can neglect the coupled Maxwell's equations and solve
the conservation equations perturbatively and analytically. After that, we will discuss the anistropic 
transverse flow induced by the inhomogenous magnetic fields. 

The paper is organized as follows. In Sec.\ref{solving}, we solve the MHD equations
with a transverse external magnetic field perturbatively by approximating the spatial dependence of magnetic fields via the Fourier decomposition and obtain
the analytic solution for each moment up to the leading-order corrections. In Sec.\ref{aniso_flow}, we then employ our solution to a concrete example and discuss the modifications of fluid velocity and energy density. Finally, 
we make conclusions and outlook in the last section.  

\section{Perturbative Solutions for Weak Magnetic Fields \label{solving}}
We consider an inviscid fluid coupled to a magnetic field $B^{\mu}$. In the flat spacetime $\eta_{\mu\nu}=\text{diag}\{-,+,+,+\}$, the general form of the energy-momentum tensor is given by \cite{Gedalin:PRE1995,Huang:2009ue}
\begin{equation}
T^{\mu\nu}=(\epsilon+p+B^{2})u^{\mu}u^{\nu}+(p+\frac{1}{2}B^{2})\eta^{\mu\nu}-B^{\mu}B^{\nu},\label{eq:EMT_01}
\end{equation}
where 
\begin{equation}
B^{2}=B^{\mu}B_{\mu},\;B^{\mu}=\frac{1}{2}\epsilon^{\mu\nu\alpha\beta}u_{\nu}F_{\alpha\beta}.
\end{equation}
Here $u^{\mu}$, $\epsilon$, and $p$ correspond to the four velocity of fluid, energy density, and pressure, respectively. Also, $\epsilon^{0123}=-\epsilon_{0123}=1$ represents the Levi-Civita tensor. In our convention, the velocity of the fluid satisfies $u^{\mu}u_{\mu}=-1$. The energy-momentum tensor should follow the conservation equations $\nabla_{\mu}T^{\mu\nu}=0$. In general, the presence of external fields may induce internal electromagnetic fields of the fluid, where the latter are dictated by Maxwell's equations. One should thus solve the conservation equations and Maxwell's equations coupled to each other. In this work, we only focus on the effects of an external magnetic fields and discard the back-reaction from the internal fields. Since the external magnetic field is generated by external sources, it can take an arbitrary form. Therefore, the energy-momentum tensor will be solely governed by the conservation equations. By implementing the projection of $\nabla_{\mu}T^{\mu\nu}=0$ along the longitudinal and transverse directions with respect to $u^{\mu}$, one can rewrite the conservation equations as  
\begin{eqnarray}\label{cons_original}\nonumber
u_{\nu}\nabla_{\mu}T^{\mu\nu}\nonumber & = & -(u\cdot\nabla)(\epsilon+\frac{1}{2}B^{2})-(\epsilon+p+B^{2})(\nabla\cdot u)-u_{\nu}\nabla_{\mu}(B^{\mu}B^{\nu})=0,
\\
\Delta_{\nu\alpha}\nabla_{\mu}T^{\mu\nu}&=& (\epsilon+p+B^{2})(u\cdot\nabla)u_{\alpha}+\Delta_{\nu\alpha}\nabla^{\nu}(p+\frac{1}{2}B^{2})-\Delta_{\nu\alpha}\nabla_{\mu}(B^{\mu}B^{\nu})=0,
\end{eqnarray}
where $\Delta^{\nu\alpha}=\eta^{\nu\alpha}+u^{\nu}u^{\alpha}$.

To simplify the problem and qualitatively delineate the practical condition in heavy ion collisions,
 we assume that the external magnetic field is perpendicular to the reaction plane, which depends
 on only one of transverse coordinates and the proper time $\tau$ for an fluid following the Bjorken
 expansion in the longitudinal direction. Nevertheless, we also present the results for a magnetic 
field depending on $\tau$ and rapidity $\eta$ in Appendix \ref{rapidity}. Here we assume that the magnitude of
 the magnetic field is suppressed by the energy density of the fluid, $B^2/\epsilon\ll 1$, which
 allows us to neglect nonlinear effects in $B^2$. Practically, such an assumption is not far from the 
scenario in heavy ion collisions, in which the magnetic field drops rapidly with respect to time \
\cite{Kharzeev:2007jp,Tuchin:2013apa}. For example, the ratio of magnetic energy density over 
the fluid energy density is $\sim 0.2$ in a typical Au-Au collisions at $\sqrt{s_{NN}}=200$GeV \cite{Roy:2015coa}. 
Although the nonlinear effects are substantial in very early
 times, in most of the time period in hydrodynamic evolution, the magnetic field could be subleading compared
 with the energy density of the fluid. Moreover, we impose the conformal invariance for the equation of state, 
which gives $p=\epsilon/3$.  
  
We now seek the perturbative solution in the presence of a weak external magnetic field pointing along the $y$ direction in an inviscid fluid following the Bjorken expansion along the $z$ direction, where $B_y$ depends on $\tau=\sqrt{t^2-z^2}$ and $x$. The setup reads
\begin{eqnarray}
{\bf B}=\lambda B_y(\tau,x)\hat{y}, \quad \epsilon=\epsilon_0(\tau)+\lambda^2 \epsilon_1(\tau,x),\quad u_{\mu}=(1,\lambda^2 u_{x}(\tau,x),0,0),
\end{eqnarray}
where $\epsilon_0(\tau)=\epsilon_c/\tau^{4/3}$. Here $\tau$ is rescaled by an initial time $\tau_0$ and $\epsilon_c$ represents the initial energy density of the medium at $\tau_0$. In the following calculations, we will implicitly rescale $x$ by $\tau_0$ as well. We introduce $\lambda$ as an expansion parameter in calculations, which will be set to unity in the end. In such setup, the conservation equations in (\ref{cons_original}) reduce to two coupled differential equations.  
Up to $\mathcal{O}(\lambda^2)$, the two differential equations are
\begin{eqnarray}\nonumber\label{two_cons_xdep}
&&
\partial_{\tau}\epsilon_1+\frac{4\epsilon_1}{3\tau}-\frac{4\epsilon_c\partial_{x}u_{x}}{3\tau^{4/3}}+B_y\partial_{\tau}B_y+\frac{B_y^2}{\tau}=0,
\\
&&
\partial_{x}\epsilon_1-\frac{4\epsilon_c\partial_{\tau}u_{x}}{\tau^{4/3}}
+\frac{4\epsilon_cu_{x}}{3\tau^{7/3}}+3B_y\partial_{x}B_y=0.
\end{eqnarray}
The combination of two equations above yields a partial differential equation solely depending on $u_{x}$,   
\begin{eqnarray}\label{equ_xdep}
\tau^2\partial_{x}^2 u_{x}-u_{x}-3\tau^2\partial^2_{\tau}u_{x}+\tau\partial_{\tau}u_{x}+
\frac{3\tau^{7/3}}{4\epsilon_c}\partial_{x}\left(B_y^2+\tau\partial_{\tau}B_y^2\right)=0.
\end{eqnarray}

Now, the solution of $u_x(\tau,x)$ depends on the explicit form of $B_y(\tau,x)$. 
Here we consider the case when the $\tau$ dependence and $x$ dependence of $B_y$ are separable. 
When $B_y=0$, (\ref{equ_xdep}) is a homogeneous partial differential equation, which can be solved by separation of variables. The general solution takes the form,
\begin{eqnarray}
u^{h}_x(\tau,x)|_{B_y=0}=\sum_m A^m_1\left( \tau^{2/3}J_{\frac{1}{3}}\left[\frac{m\tau}{\sqrt{3}}\right]
+A^m_2\tau^{2/3}Y_{\frac{1}{3}}\left[\frac{m\tau}{\sqrt{3}}\right]\right)\left[\cos(mx)+A^m_3\sin(mx)\right],
\end{eqnarray}
where $m$ can be real or imaginary numbers and $A^m_{1,2,3}$ are integration constants.   
To find the solution for $B_y\neq 0$, we may rewrite $B_y^2$ into a Fourier series on the bases of the $x$-dependence part of the general solution. This is the key step to convert the solution of a partial differential equation into the summation of solutions of ordinary differential equations although the trick is only valid for a finite region where the spatial part of $B_y$ can be accurately approximated by the Fourier series.
Since the magnetic field generated in peripheral heavy ion collisions should be even with respect to $x$ and most prominent at $x=0$ on the transverse plane, we may decompose it into a cosine series
\begin{eqnarray}
B^2_y(\tau,x)=\sum_k \tilde{B}_k^2(\tau)\cos(kx), 
\end{eqnarray}
where $k\geq 0$ are now real integers.
For simplicity, we may further assume that $\tilde{B}^2_k(\tau)=\tau^n B_k^2$ for $n<0$ with $B_k$ being constants, which approximately characterizes the decay of magnetic fields in heavy ion collisions.   
Accordingly, we make the following ansatz,
\begin{eqnarray}
u_{x}(\tau,x)=\sum_m \left[a_m(\tau)\cos(m x)+b_m(\tau)\sin(m x)\right],
\end{eqnarray}
and solve (\ref{equ_xdep}). Note that $u_x(\tau,x)=0$ when $k=0$ \footnote{Although one could find a trivial solution, $u_x(\tau)=C \tau^{1/3}$, from the second equation of (\ref{two_cons_xdep}) when $\partial_xB_y=0$, one should set $C=0$ since the solution is irrelevant to the magnetic field.}, which is also shown in \cite{Roy:2015kma} in the absence of the spatial dependence of magnetic fields. For each moment with $k>0$, we find $m=k$ and $a_m(\tau)=0$, while $b_k(\tau)$ is solved from the following ordinary differential equation,
\begin{eqnarray}\label{eq_bk}
(3\tau^2\partial_{\tau}^2-\tau\partial_{\tau}+k^2\tau^2+1)b_k(\tau)+\frac{3B_k^2}{4\epsilon_c}k(n+1)\tau^{n+7/3}=0.
\end{eqnarray}
After solving (\ref{eq_bk}) analytically, the perturbative solution turns out to be 
\begin{eqnarray}\label{ux_sol}
u_x(\tau,x)=u^h_x(\tau,x)+u^{ih}_x(\tau,x),
\end{eqnarray}
where 
\begin{eqnarray}
u^{h}_x(\tau,x)=\sum_{k\neq 0}\left( C^k_1\tau^{2/3}J_{\frac{1}{3}}\left[\frac{k\tau}{\sqrt{3}}\right]
+C^k_2\tau^{2/3}Y_{\frac{1}{3}}\left[\frac{k\tau}{\sqrt{3}}\right]\right)\sin(kx)
\end{eqnarray}
and
\begin{eqnarray}\nonumber
u^{ih}_x(\tau,x)&=&\sum_{k\neq 0}
\frac{3B_k^2}{8\epsilon_c} k (1+n) \tau ^{\frac{7}{3}+n} \Bigg(-\frac{3}{4+3 n} \textrm{}_0F_1
\left[\frac{4}{3},-\frac{1}{12} k^2 \tau ^2\right]\mbox{} _pF_q
\left[\left\{\frac{2}{3}+\frac{n}{2}\right\},\left\{\frac{2}{3},\frac{5}{3}+\frac{n}{2}\right\},
-\frac{1}{12} k^2 \tau ^2\right]
\\
&&+\frac{1}{2+n} \textrm{}_0F_1\left[\frac{2}{3},-\frac{1}{12} k^2 \tau ^2\right] \mbox{}_pF_q\left[\left\{1+\frac{n}{2}\right\},\left\{\frac{4}{3},2+\frac{n}{2}\right\},-\frac{1}{12} k^2 \tau ^2\right]\Bigg)\sin(k x)
\end{eqnarray}
correspond to the homogeneous and inhomogeneous solutions, respectively. Here $J_{\frac{1}{3}}$ and $Y_{\frac{1}{3}}$ are Bessel functions, while $_0F_1$ and $_pF_q$ are hypergeometric functions. For the homogeneous solution, there exist two integration constants for each $k$, which are usually determined by initial conditions at $\tau=\tau_0$. Nevertheless, we may fix $C^k_1$ and $C^k_2$ by introducing an initial condition at late time when $\tau\rightarrow \infty$. Numerically, one should solve (\ref{eq_bk}) inversely in time since the late-time dynamics is simply governed by ideal hydrodynamics but the early-time condition is unknown. 
Such an initial condition at $\tau\rightarrow \infty$ can be derived from making a serial expansion of (\ref{eq_bk}) in large $\tau$ and solving for the asymptotic solution order by order. Alternatively, we will obtain the same condition in the following by imposing the regularity condition on the analytic solution of $u_x$ at $\tau\rightarrow \infty$. 

We expect $u_x(\infty,x)\rightarrow 0$ since $B_y^2(\infty,x)\rightarrow 0$. By making late-time expansion of $u_x$, one finds that both $u^h_x$ and $u^{ih}_x$ take the asymptotic form as $\omega(\tau)\tau^{1/6}$ with $\omega(\tau)$ denotes an oscillatory function. It turns out that the proper choice which leads to the cancellation of $\tau^{1/6}$ divergence reads
\begin{eqnarray}\nonumber
C^k_1&=&\frac{B_k^2 2^{-\frac{7}{3}+n} 3^{\frac{1}{3}+\frac{n}{2}} k^{-\frac{2}{3}-n} (1+n) \pi ^2 }{\epsilon_c \Gamma\left[\frac{1}{3}-\frac{n}{2}\right] \Gamma\left[-\frac{n}{2}\right]} 
\left(\csc\left[\frac{n \pi }{2}\right]+2  \sec\left[\frac{1}{6} (\pi +3 n \pi )\right]\right),
\\
C^k_2&=&\frac{B_k^22^{-\frac{7}{3}+n} 3^{\frac{5}{6}+\frac{n}{2}} k^{-\frac{2}{3}-n} (1+n) \pi }{\epsilon_c \Gamma\left[\frac{1}{3}-\frac{n}{2}\right]}  \Gamma\left[1+\frac{n}{2}\right].
\end{eqnarray}
When $\tau\rightarrow \infty$, such a choice of $C^k_1$ and $C^k_2$ yields 
\begin{eqnarray}\label{asymptotic_ux}
u_x(\tau,x)\rightarrow -\sum_{k\neq 0}\frac{3 B_k^2 (1+n)}{4 \epsilon_c k}\tau^{n+1/3}\sin(kx),
\end{eqnarray}
which is consistent with the asymptotic solution of (\ref{equ_xdep}) (or (\ref{eq_bk})) obtained from the serial expansion in $\tau$ as the initial condition.   
After solving $u_x(\tau,x)$, we can derive the corresponding energy-density modification from the second equation of (\ref{two_cons_xdep}), which is given by
\begin{eqnarray}\label{solve_e1}
\epsilon_1(\tau,x)=-\frac{3 B_0^2 (2+n) \tau^n}{8+6 n}
-\sum_{k\neq 0}\frac{\cos(kx)}{k}\left[\frac{4\epsilon_c\partial_{\tau}b_k(\tau)}{\tau^{4/3}}-\frac{4\epsilon_cb_k(\tau)}{3\tau^{7/3}}+\frac{3k}{2}\tilde{B}^2_k
(\tau)\right].
\end{eqnarray} 
The result reads
\begin{eqnarray}\label{E1_sol}
\epsilon_1(\tau,x)=-\frac{3 B_0^2 (2+n) \tau^n}{8+6 n}+\epsilon^h_1(\tau,x)+\epsilon^{ih}_1(\tau,x),
\end{eqnarray}
where
\begin{eqnarray}\nonumber
\epsilon^h_1(\tau,x)&=& \sum_{k\neq 0}\frac{B_k^2}{k^{\frac{2}{3}+n}}  \frac{3^{-\frac{2}{3}+
\frac{n}{2}}}{2^{\frac{1}{3}-n}}  (1+n) \pi^2 \left\{
3Y_{-\frac{2}{3}}\left[\frac{k \tau }{\sqrt{3}}\right] 
\csc\left[\frac{n \pi }{2}\right]-\sqrt{3} J_{-\frac{2}{3}}\left[\frac{k \tau }{\sqrt{3}}\right] 
\left(\csc\left[\frac{n \pi }{2}\right]+2 \sec\left[\frac{1}{6} (\pi +3 n \pi )\right]\right)\right\}
\\
&&\times\left(\tau^{2/3}\Gamma\left[\frac{1}{3}-\frac{n}{2}\right] \Gamma\left[-\frac{n}{2}\right]\right)^{-1} \cos(kx)
\end{eqnarray}
and
\begin{eqnarray}\nonumber
\epsilon^{ih}_1(\tau,x)&=&\sum_{k\neq 0}\frac{3B_k^2 \tau ^n}{8} 
\Bigg(\frac{1+n}{8+6 n} \left(16 _0F_1\left[\frac{4}{3},-\frac{1}{12} k^2 \tau ^2\right]
-3 k^2 \tau ^2 \mbox{}_0F_1\left[\frac{7}{3},-\frac{1}{12} k^2 \tau ^2\right]\right) 
\\
&&\times \mbox{}_pF_q\left[\left\{\frac{2}{3}+\frac{n}{2}\right\},\left\{\frac{2}{3},\frac{5}{3}
+\frac{n}{2}\right\},-\frac{1}{12} k^2 \tau ^2\right] \\\nonumber
\\
&&+\frac{k^2 (1+n) \tau ^2}{2+n} \mbox{}_0F_1\left[\frac{5}{3},-\frac{1}{12} k^2 \tau ^2\right]
 \mbox{}_pF_q\left[\left\{1+\frac{n}{2}\right\},\left\{\frac{4}{3},2+\frac{n}{2}\right\},
-\frac{1}{12} k^2 \tau ^2\right]-4\Bigg) \cos(kx).
\end{eqnarray}
When $\tau\rightarrow\infty$, from (\ref{asymptotic_ux}) and (\ref{solve_e1}), one immediately finds that the asymptotic form of the energy-density correction becomes
\begin{eqnarray}
\epsilon_1(\tau,x)\rightarrow -\frac{3 B_0^2 (2+n) \tau^n}{8+6 n}-\sum_{k\neq 0}\frac{3B_k^2\tau^n}{2}\cos(kx).
\end{eqnarray}
From the analytic expression of each moment of $u_x(\tau, x)$ in (\ref{ux_sol}), one immediately notes that $u_x(\tau,x)=0$ when $n=-1$, which can also be found from (\ref{equ_xdep}), where the inhomogeneous term vanishes in such a case. For $B_y^2(\tau,x)=\tilde{B}_y^2(x)\tau^{-1}$, it turns out that the solution of (\ref{two_cons_xdep}) simply reduces to 
\begin{eqnarray}
\epsilon_1(\tau,x)=-\frac{3\tilde{B}_y^2(x)}{2\tau}=-\frac{3B_y^2(\tau,x)}{2},\quad u_x(\tau,x)=0
\end{eqnarray}
where the $x$-dependent part of the magnetic field $\tilde{B}_y(x)$ is arbitrary.  
 
Note that the solutions above in (\ref{ux_sol}) and (\ref{E1_sol}) for $u_x$ and $\epsilon_1$ are invalid for $n=-4/3$ and $n=-2$, which can be observed from the divergence in the inhomogeneous solution. In the former case for $n=-4/3$, the ratio $B_y^2(\tau,x)/\epsilon(\tau)$ up to $\lambda^2$ becomes constant in $\tau$. Even in the space-independent solution for $k=0$, the modification of the energy density has a logarithmic correction on top of the power-law decay with respect to time, which is distinct from the general pattern for other exponents $n$ \cite{Roy:2015kma}. The latter case for $n=-2$ corresponds to the ideal magnetohydrodynamics, where the magnetic field satisfies the "frozen-flux condition",
\begin{equation}
(u\cdot\nabla)\left(\frac{B^{\mu}}{s}\right)=\frac{1}{s}\left[(B\cdot\nabla)u^{\mu}+u^{\mu}\nabla\cdot B\right]\label{eq:frozen_flux_01},
\end{equation} 
which stems from the Maxwell's equations and conservation of the entropy-density current. 
One could check that our setup satisfies such a condition up to $\mathcal{O}(\lambda^2)$, where
 the exponent of power-law decay in proper time for the magnetic field is now same as the one for entropy 
density $s$ at $\mathcal{O}(\lambda^2)$. For the space-independent solution, the energy density is
 unmodified under this condition \cite{Roy:2015kma}. The solutions satisfying the initial condition in (\ref{asymptotic_ux}) for these two particular cases are shown in the following. For $n=-4/3$, 
\begin{eqnarray}
\nonumber
u_x(\tau,x)&=&\sum_{k\neq 0}\frac{B_k^2\sin(kx) 
\pi  (k \tau )^{2/3}}{144\times 2^{2/3} 3^{5/6} \epsilon_c 
\Gamma\left[\frac{4}{3}\right]^2} 
\Bigg\{ \Gamma\left[\frac{1}{3}\right]\Bigg(-2^{1/3} 3^{7/6} (k \tau )^{2/3} 
\mbox{}_pF_q\left[\left\{\frac{1}{3}\right\},\left\{\frac{4}{3},\frac{4}{3}\right\},\frac{-k^2\tau^2}{12} \right]
\\\nonumber
&&+2\sqrt{3} \Gamma\left[\frac{4}{3}\right] \Gamma\left[\frac{1}{3}\right]\Bigg)\left(J_{\frac{1}{3}}\left[\frac{k \tau }{\sqrt{3}}\right]-\sqrt{3} Y_{\frac{1}{3}}\left[\frac{k \tau }{\sqrt{3}}\right]\right)-12 \sqrt{3} \Gamma\left[\frac{4}{3}\right]^2J_{\frac{1}{3}}\left[\frac{k \tau }{\sqrt{3}}\right] 
G^{20}_{13}\left[\frac{k^2\tau^2}{12}\Big|^1_{0,0,1/3}\right] \Bigg\},
\\\nonumber
\epsilon_1(\tau,x)&=&\sum_{k\neq 0}\frac{-B_k^2\cos(kx)}
{
1944 \left(k \tau ^5\right)^{1/3} \Gamma\left[\frac{4}{3}\right]^2}
\left\{
2^{2/3} 3^{19/6} \pi k \tau  \left( 3^{1/6} (k \tau )^{2/3} Y_{-\frac{2}{3}}\left[\frac{k \tau }{\sqrt{3}}\right] \Gamma\left[\frac{1}{3}\right]-2^{2/3} \mbox{}_0F_1\left[\frac{1}{3},-\frac{1}{12} k^2 \tau ^2\right]\right) 
\right.
\\\nonumber
&&\times\mbox{}_pF_q\left[\left\{\frac{1}{3}\right\},\left\{\frac{4}{3},\frac{4}{3}\right\},
-\frac{1}{12} k^2 \tau ^2\right]+1080 \Gamma\left[\frac{4}{3}\right] 
\Gamma\left[-\frac{5}{3}\right](k \tau )^{1/3} - 2^{10/3} 3^{1/6} 
\Gamma\left[\frac{4}{3}\right]\Gamma\left[-\frac{2}{3}\right]^2 \pi  k\tau  
\\
&&\left. \times\Bigg(\sqrt{3} Y_{-\frac{2}{3}}\left[\frac{k \tau }{\sqrt{3}}\right]
 -J_{-\frac{2}{3}}\left[\frac{k \tau }{\sqrt{3}}\right]\Bigg) - 2^{7/3} 3^{19/6}
 k\pi\tau \Gamma\left[\frac{4}{3}\right]^2J_{-\frac{2}{3}}\left[\frac{k \tau }{\sqrt{3}}\right] 
G^{20}_{13}\left[\frac{k^2\tau^2}{12}\Big|^1_{0,0,1/3}\right]
\right\}
-\frac{B_0^2\log\tau}{3\tau^{4/3}},
\end{eqnarray}
where 
\begin{eqnarray}
G^{mn}_{pq}\left[z\Big|^{a_1,\cdots, a_p}_{b_1,\cdots, b_q}\right]=
\frac{1}{2\pi i}\int \frac{\Gamma \left(1-a_1-s\right) \ldots  \Gamma \left(1-a_n-s\right)\Gamma \left(b_1+s\right) \ldots  \Gamma \left(b_m+s\right)}{\Gamma \left(a_{n+1}+s\right) \ldots  \Gamma \left(a_p+s\right)\Gamma \left(1-b_{m+1}-s\right) \ldots  \Gamma \left(1-b_q-s\right)}z^{-s}ds
\end{eqnarray}
is the Mijer G function. 
For $n=-2$, 
\begin{eqnarray}\nonumber
u_x(\tau,x)&=&\sum_k\frac{B_k^2\pi\sin(kx)}{48\times 6^{2/3} \epsilon_c \Gamma\left[\frac{2}{3}\right]^2}
\left\{J_{\frac{1}{3}}\left[\frac{k \tau }{\sqrt{3}}\right] \Gamma\left[-\frac{1}{3}\right] \Bigg(2^2 3^{4/3} k^{2/3}\mbox{}_pF_q\left[\left\{-\frac{1}{3}\right\},\left\{\frac{2}{3},\frac{2}{3}\right\},-\frac{1}{12} k^2 \tau ^2\right]
\right.
\\\nonumber
&& \left.
+2^{4/3} (k^{2}\tau)^{2/3}\Gamma\left[\frac{2}{3}\right]\Gamma\left[-\frac{1}{3}\right]\Bigg)
+2^{1/3} 3\Gamma\left[\frac{2}{3}\right]^2  \left(k^2 \tau \right)^{2/3} \left(J_{\frac{1}{3}}
\left[\frac{k \tau }{\sqrt{3}}\right]-\sqrt{3} Y_{\frac{1}{3}}\left[\frac{k \tau }{\sqrt{3}}\right]\right) 
 G^{20}_{13}\left[\frac{k^2\tau^2}{12}\Big|^1_{0,0,-1/3}\right]
\right\},
\\\nonumber
\epsilon_1(\tau,x)&=&-\sum_k\frac{B_k^2\cos(kx)}{432 \tau ^2 \Gamma\left[\frac{2}{3}\right]^2}
\left\{\pi  (k \tau )^{2/3} J_{-\frac{2}{3}}\left[\frac{k \tau }{\sqrt{3}}\right] \Gamma\left[-\frac{1}{3}\right] \Bigg(2^{10/3} 3^{13/6} \mbox{}_pF_q\left[\left\{-\frac{1}{3}\right\},\left\{\frac{2}{3},\frac{2}{3}\right\},-\frac{1}{12} k^2 \tau ^2\right]
\right.
\\\nonumber
&&
+2^{8/3}3^{5/6}\Gamma\left[\frac{2}{3}\right]\Gamma\left[-\frac{1}{3}\right]
(k\tau)^{2/3}\Bigg)-216\Gamma\left[\frac{2}{3}\right]\Gamma\left[-\frac{1}{3}\right]
\\
&& \left.
+2^{5/3} 3^{4/3} \pi  (k \tau )^{4/3} \Gamma\left[\frac{2}{3}\right]^2 \left(\sqrt{3} J_{-\frac{2}{3}}
\left[\frac{k \tau }{\sqrt{3}}\right]-3 Y_{-\frac{2}{3}}\left[\frac{k \tau }{\sqrt{3}}\right]\right) 
G^{20}_{13}\left[\frac{k^2\tau^2}{12}\Big|^1_{0,0,-1/3}\right]\Bigg) \right\}.
\end{eqnarray}

Finally, we mention the validity of our perturbative solution. Recall that our result is the leading-order solution based on the constraint $B(\tau,x)^2/\epsilon_0(\tau)< 1$. Our setup implies $\sum_k B_k^2\tau^{n+4/3}/\epsilon_c< 1$ with $\sum_k B_k^2/\epsilon_c<1$. When $n\leq -4/3$, the perturbative expansion is legitimate for arbitrary late times $\tau\geq 1$. Nonetheless, when $n>-4/3$, the perturbative solution is only valid withing the time period   
\begin{eqnarray}
1\leq \tau<\left(\sum_k\epsilon_cB_k^{-2}\right)^{1/(n+4/3)}.
\end{eqnarray}


\section{Anisotropic flow from magnetic fields \label{aniso_flow}}             
\begin{figure}[t]
		{\includegraphics[width=8cm,height=6.5cm,clip]{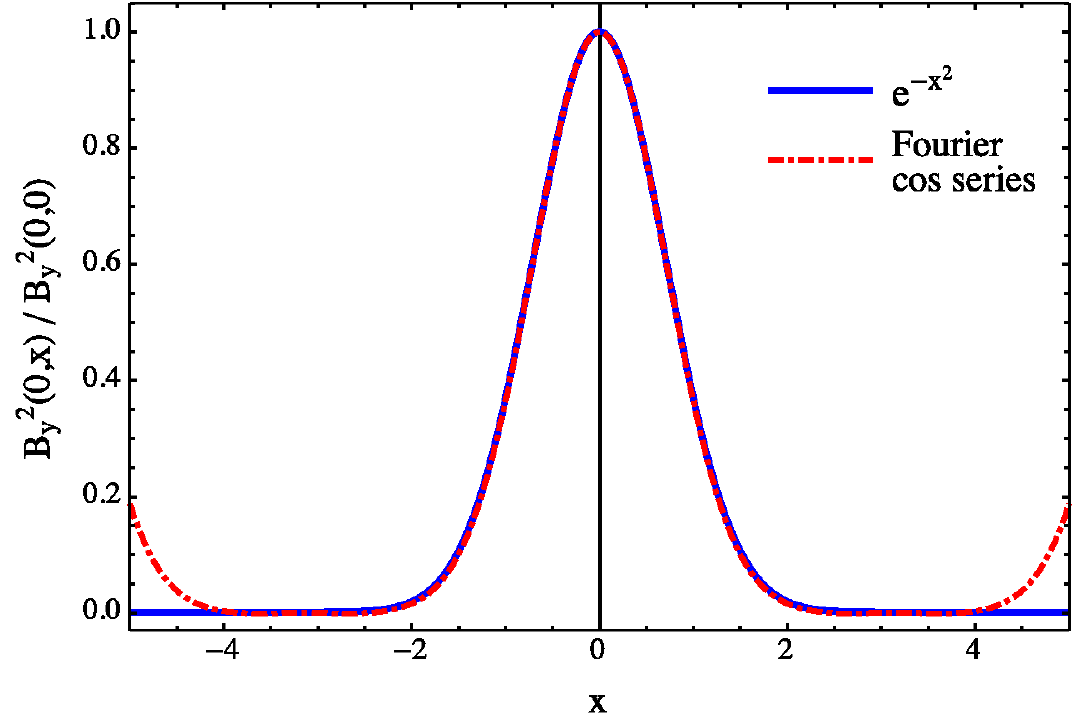}}
		\caption{A comparison between the approximated $B_y$ in Fourier cosine series (red-dashed) and the genuine $B_y$ in a Gaussian distribution (blue).}\label{B_field_plot}
\end{figure}

In order to gain some phenomenological insights from the perturbative solutions, we consider the following profile of the magnetic field,
\begin{eqnarray}
{\bf B}=B_y(\tau,x)\hat{y}=B_c\tau^{n/2}e^{-x^2/2}\hat{y},
\end{eqnarray}
where we approximate the spatial dependence of the magnetic field as a Gaussian distribution. Recall that all spacetime coordinates are rescaled by the initial time $\tau_0$. Here we further choose the spatial width of the magnetic field about the same size as $\tau_0$, which allows us to reproduce $B_y(\tau,x)$ via the Fourier expansion with finite leading moments. 
Explicitly, we approximate
\begin{eqnarray}
B_y(\tau,x)^2=B_c^2\tau^n\left(0.28, +0.44 \cos x+0.21 \cos 2x+0.06 \cos 3x+0.01 \cos 4x\right),
\end{eqnarray}         
which reproduce the Gaussian distribution within $-\pi<x<\pi$, where $0.28B_c^2,\cdots, 0.01B_c^2$ as the Fourier coefficients correspond to $B_0^2,\cdots,B_4^2$. Although the mismatches emerge at larger $|x|$ as illustrated in Fig.\ref{B_field_plot} due to the oscillatory property of cosine functions, the magnetic field almost reduces to zero in the fringes for $|x|\approx \pi$. Consequently, we only have to focus on the valid region for $-\pi<x<\pi$. Although the interpolation near the fringes could be nontrivial, the prominent effects led by magnetic fields are captured by the perturbative solutions in the central region. In general, the spatial width of the magnetic field could be larger than $\tau_0$.
 In practice, the spatial width depends of $B_y$ depends on the impact parameters of peripheral collisions. 
Technically, one has to rescale $x$ by the spatial width to construct the Fourier series of the Gaussian 
distribution, where we elaborate the details of rescaling in Appendix \ref{rescaling}.

\begin{figure}[t]
	\begin{minipage}{8cm}
			{\includegraphics[width=8cm,height=6.5cm,clip]{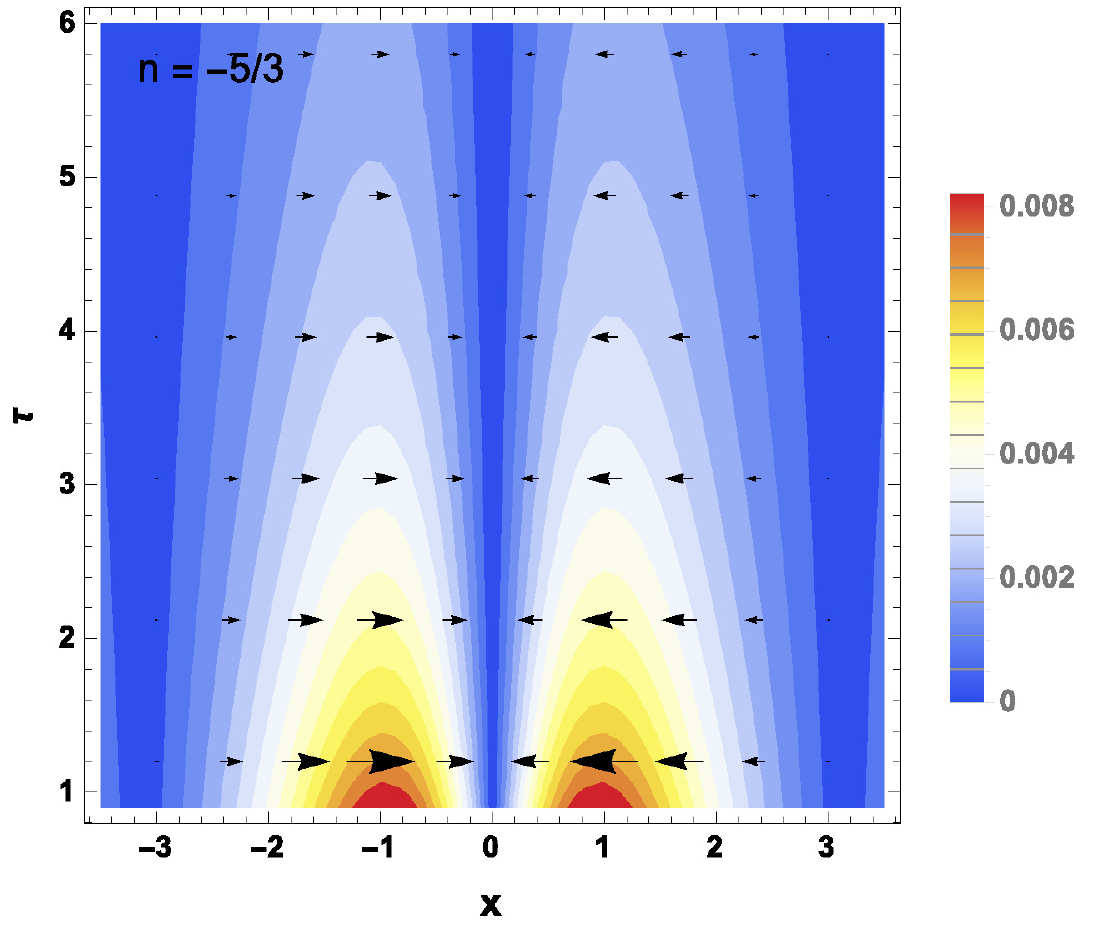}}
			\caption{Velocity plot for $v_x(\tau,x)=u^x/u^{\tau}$ with $n=-5/3$. Here the horizontal axis and vertical axis correspond to $x$ and $\tau$, respectively. The background colors represent the magnitudes of $v_x$.}\label{v_plot_x_tau}
	\end{minipage}
	\hspace {0.5cm}
	\begin{minipage}{8cm}
			{\includegraphics[width=8cm,height=6.5cm,clip]{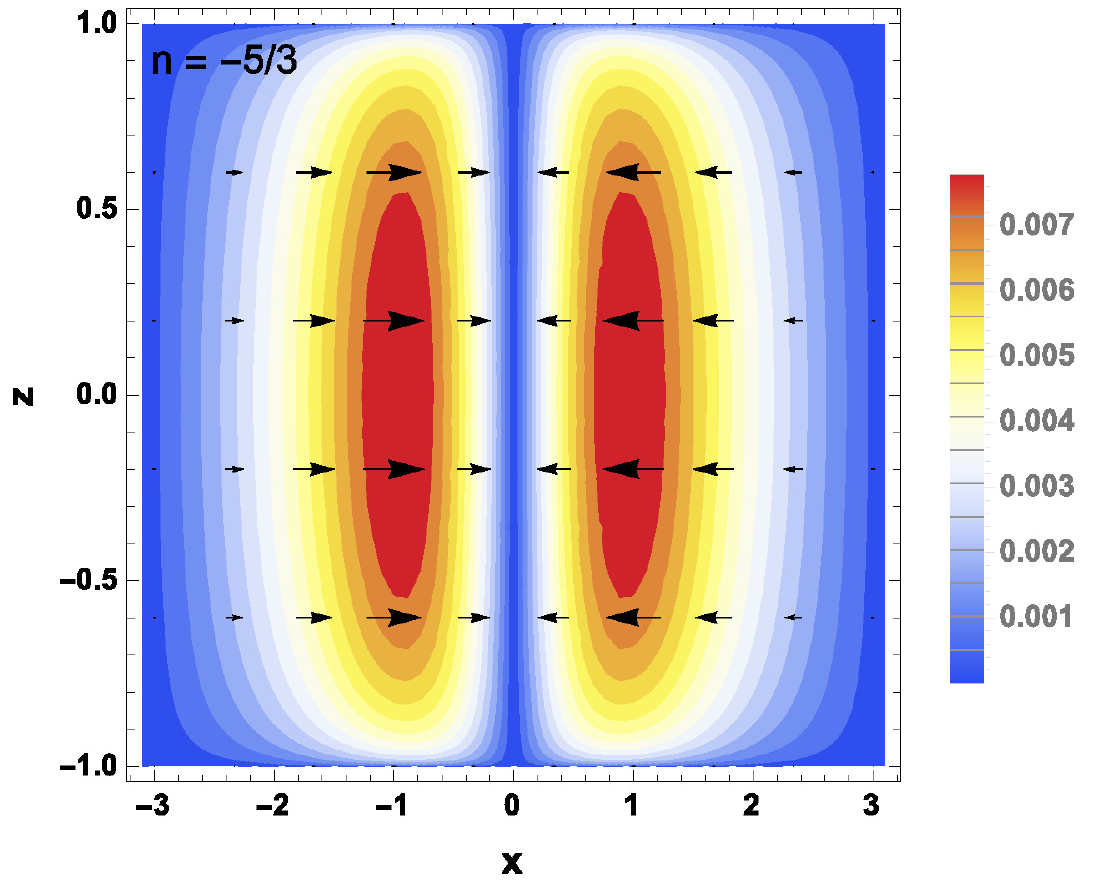}}
			\caption{Velocity plot for $v_x(x,z)=u^x/u^{\tau}$ at $t=1$ with $n=-5/3$. Here the horizontal axis and vertical axis correspond to $x$ and $z$, respectively. The background colors represent the magnitudes of $v_x$.}
			\label{v_plot_x_z}
	\end{minipage}
\end{figure}

\begin{figure}[t]
	\begin{minipage}{8cm}
			{\includegraphics[width=8cm,height=6.5cm,clip]{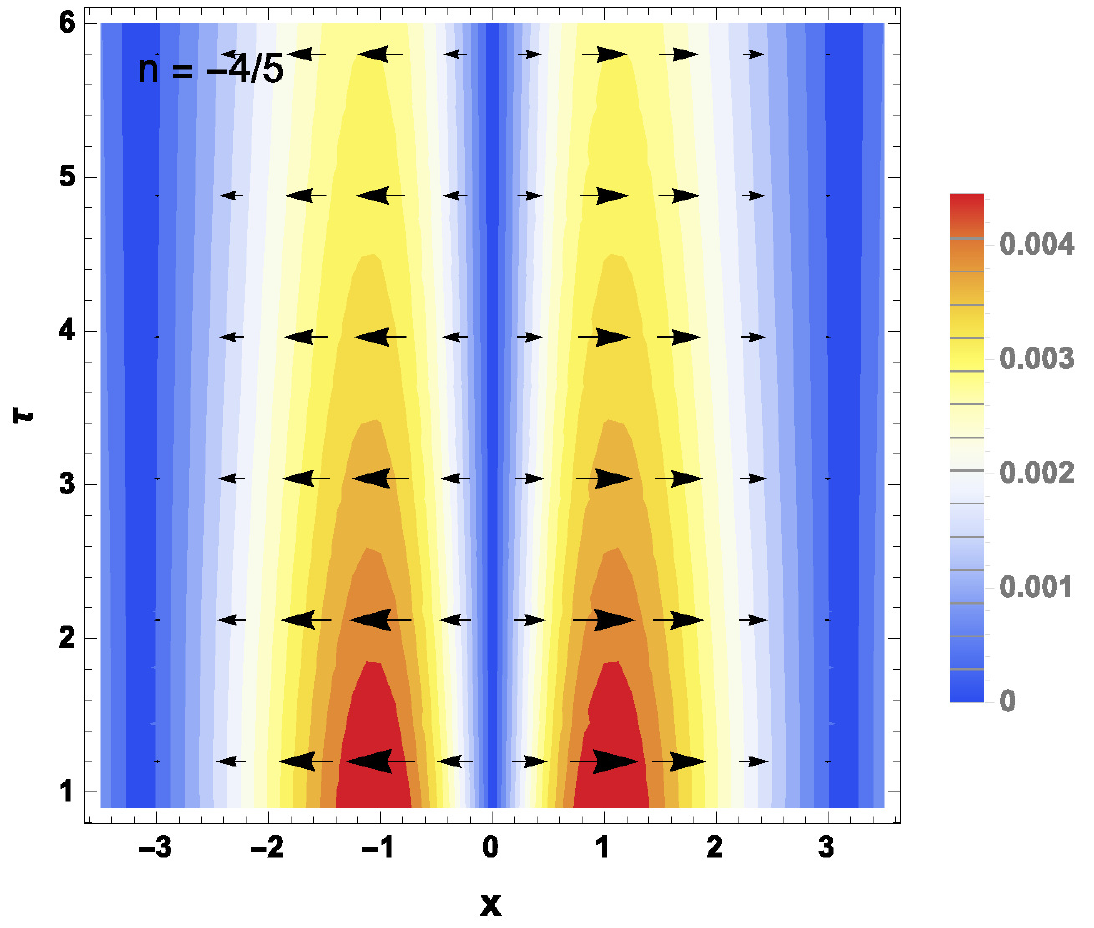}}
			\caption{Velocity plot for $v_x(\tau,x)=u^x/u^{\tau}$ with $n=-4/5$. Here the horizontal axis and vertical axis correspond to $x$ and $\tau$, respectively. The background colors represent the magnitudes of $v_x$.}\label{v_plot_x_tau2}
	\end{minipage}
	\hspace {0.5cm}
	\begin{minipage}{8cm}
			{\includegraphics[width=8cm,height=6.5cm,clip]{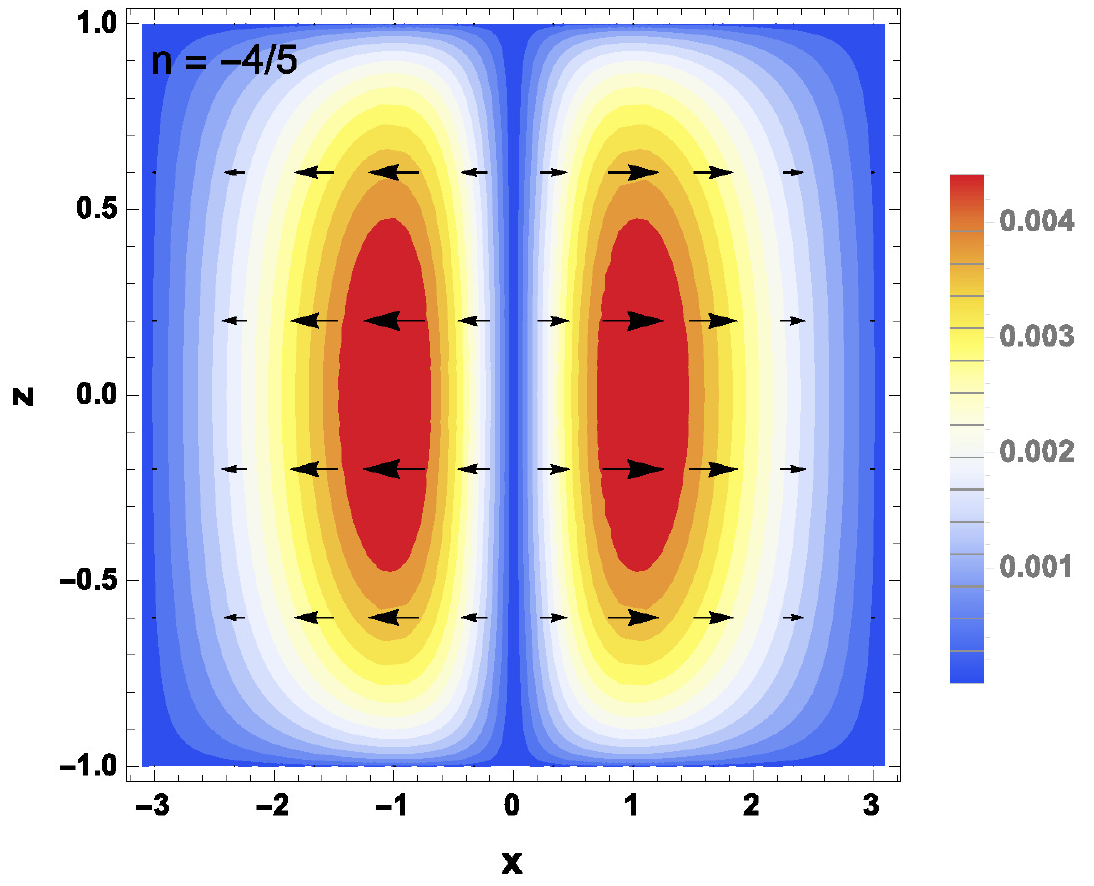}}
			\caption{Velocity plot for $v_x(x,z)=u^x/u^{\tau}$ at $t=1$ with $n=-4/5$. Here the horizontal axis and vertical axis correspond to $x$ and $z$, respectively. The background colors represent the magnitudes of $v_x$.}
			\label{v_plot_x_z2}
	\end{minipage}
\end{figure}

\begin{figure}[t]
	\begin{minipage}{8cm}
			{\includegraphics[width=7.5cm,height=5cm,clip]{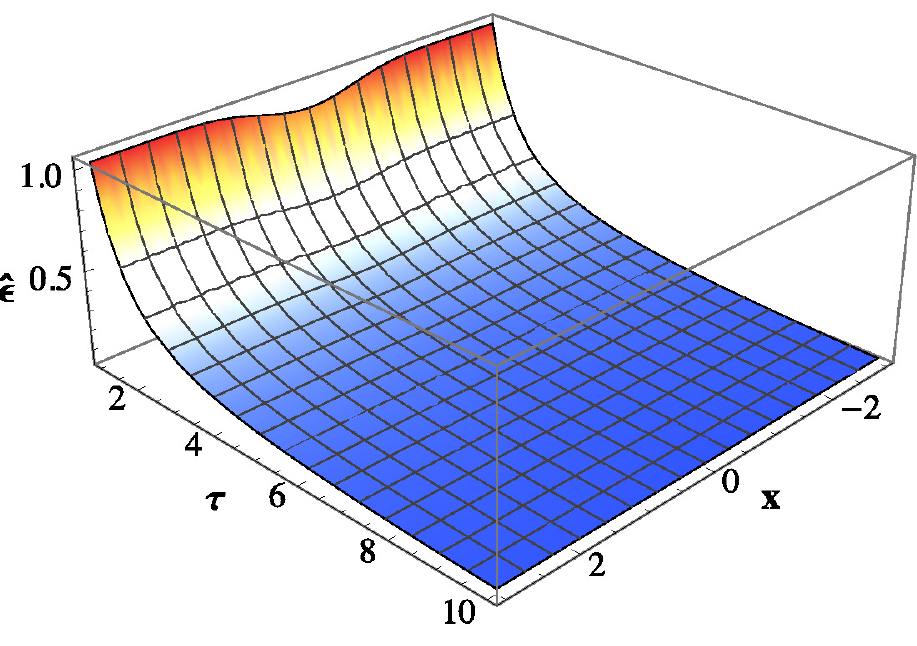}}
			\caption{The energy density ratio $\hat{\epsilon}=\epsilon/\epsilon_c$ versus $x$ and $\tau$ for $n=-5/3$.}\label{E1_plot}
	\end{minipage}
	\hspace {0.5cm}
	\begin{minipage}{8cm}
			{\includegraphics[width=7.5cm,height=5cm,clip]{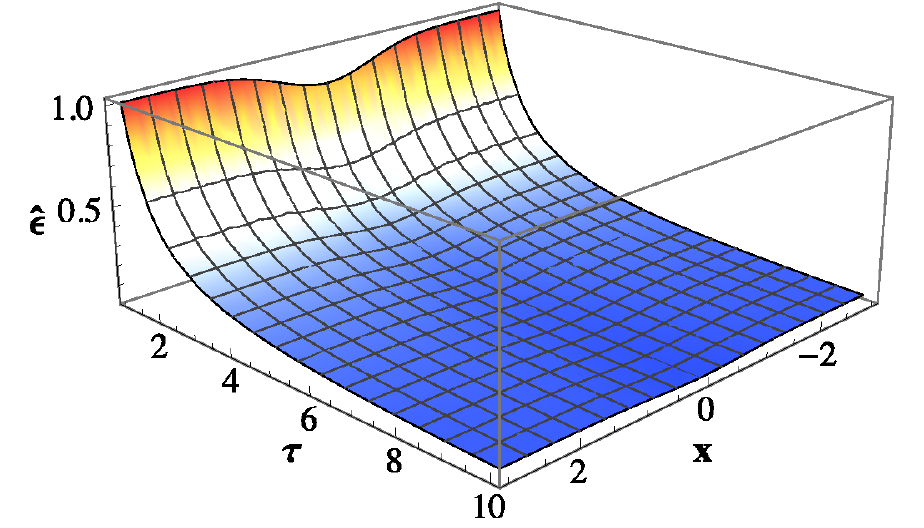}}
			\caption{The energy density ratio $\hat{\epsilon}=\epsilon/\epsilon_c$ versus $x$ and $\tau$ for $n=-4/5$.}
			\label{E1_plot2}
	\end{minipage}
\end{figure}

We choose $n=-5/3$ and $n=-4/5$ as two examples for comparisons.  
Here we make plots for fluid velocity and energy density modified by magnetic fields 
with $B_c^2/\epsilon_c=0.1$. From Fig.\ref{v_plot_x_tau} and Fig.\ref{v_plot_x_z} with $n=-5/3$, 
one finds that the magnetic field yields transverse flow pointing inward into the medium, 
where the longitudinal flow solely dictated by the Bjorken expansion is not shown here. 
From Fig.\ref{v_plot_x_tau}, one finds that the magnitude of such transverse flow gradually 
decreases with respect to proper time as expected. In addition, one observes $v_x(\tau,0)\approx 0$ 
since the fluid velocity is only modified by the spatial gradient of magnetic fields. 
Moreover, the transverse flow is more prominent in the central region with respect to the longitudinal direction
as shown in Fig.\ref{v_plot_x_z}. On the contrary, as shown in Fig.\ref{v_plot_x_tau2} 
and Fig.\ref{v_plot_x_z2} with $n=-4/5$, the magnetic field results in the transverse flow
 pointing outward, whereas other patterns regarding the magnitude of the flow are
 qualitatively in accordance with the case for $n=-5/3$. Also, as illustrated in 
Fig.\ref{E1_plot} and Fig.\ref{E1_plot2}, the energy density is most significantly 
modified in the central region $x\approx 0$, where the reduction is observed in both cases. 
The change of the direction of the transverse velocity may be anticipated since $v_x(\tau,x)=0$ 
for $n=-1$. In conclusion, the transverse flow led by a Gaussian magnetic field point 
inward for $n<-1$ and outward for $n>-1$, respectively.

\begin{figure}[t]
	\begin{minipage}{8cm}
			{\includegraphics[width=7.5cm,height=5cm,clip]{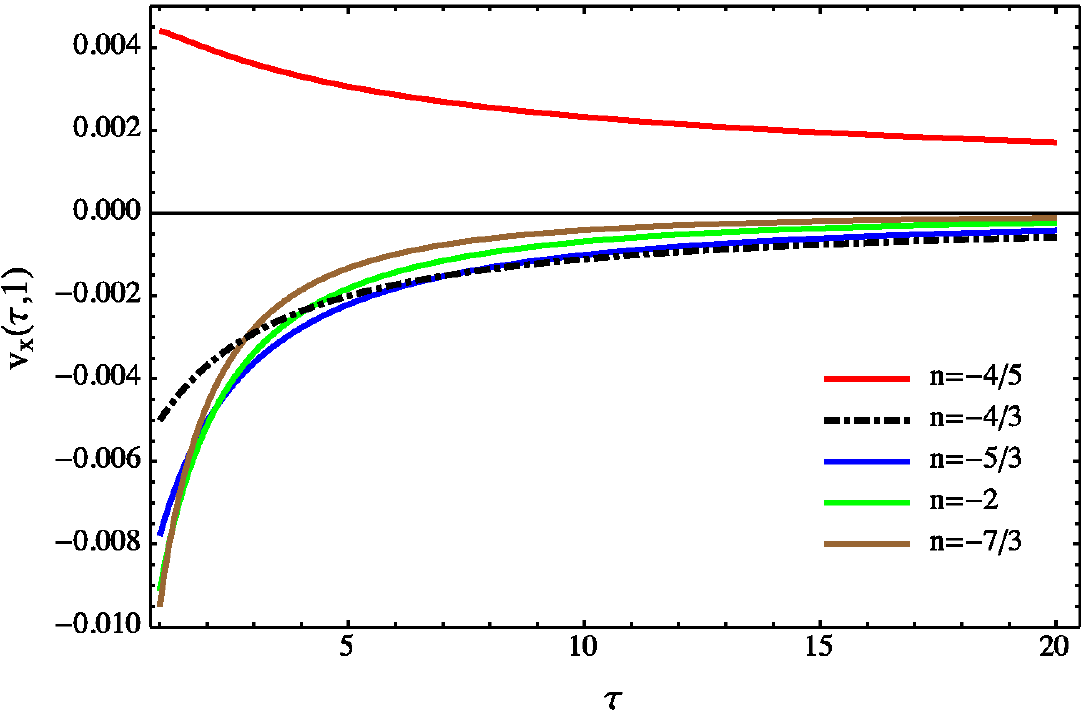}}
			\caption{$v_x$ v.s. $\tau$ plot at $x=1$ with different values of $n$. The solid curves from bottom to top at $\tau=5$ correspond to $n=-5/3$(blue), $n=-2$(green), $n=-7/3$(brown), and $n=-4/5$(red), respectively. The dashed curve corresponds to $n=-4/3$(black).}\label{v_tau_plot_diffn}
	\end{minipage}
	\hspace {0.5cm}
	\begin{minipage}{8cm}
			{\includegraphics[width=7.5cm,height=5cm,clip]{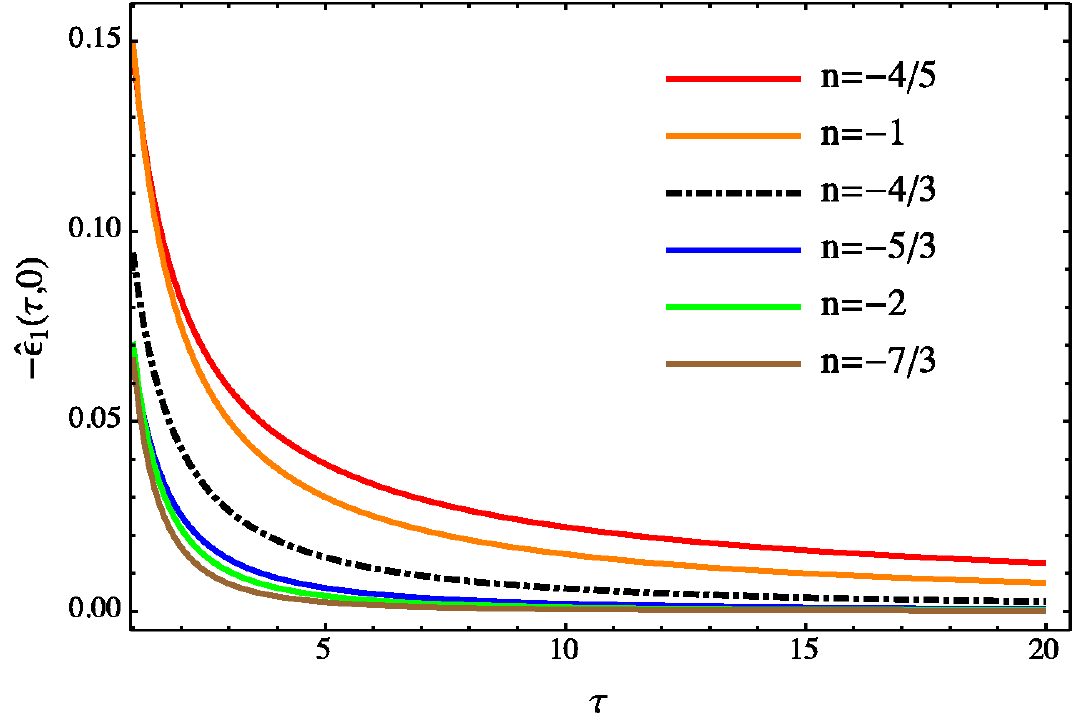}}
			\caption{$\hat{\epsilon}_1=\epsilon_1/\epsilon_c $ v.s. $\tau$ plot at $x=0$ with different values of $n$. The solid curves from bottom to top at $\tau=5$ correspond to $n=-7/3$(brown), $n=-2$(green), $n=-5/3$(blue), $n=-1$(orange), and $n=-4/5$(red), respectively. The dashed curve corresponds to $n=-4/3$(black).}
			\label{E1_tau_plot_diffn}
	\end{minipage}
\end{figure}

\begin{figure}[t]
	\begin{minipage}{8cm}
			{\includegraphics[width=7.5cm,height=5cm,clip]{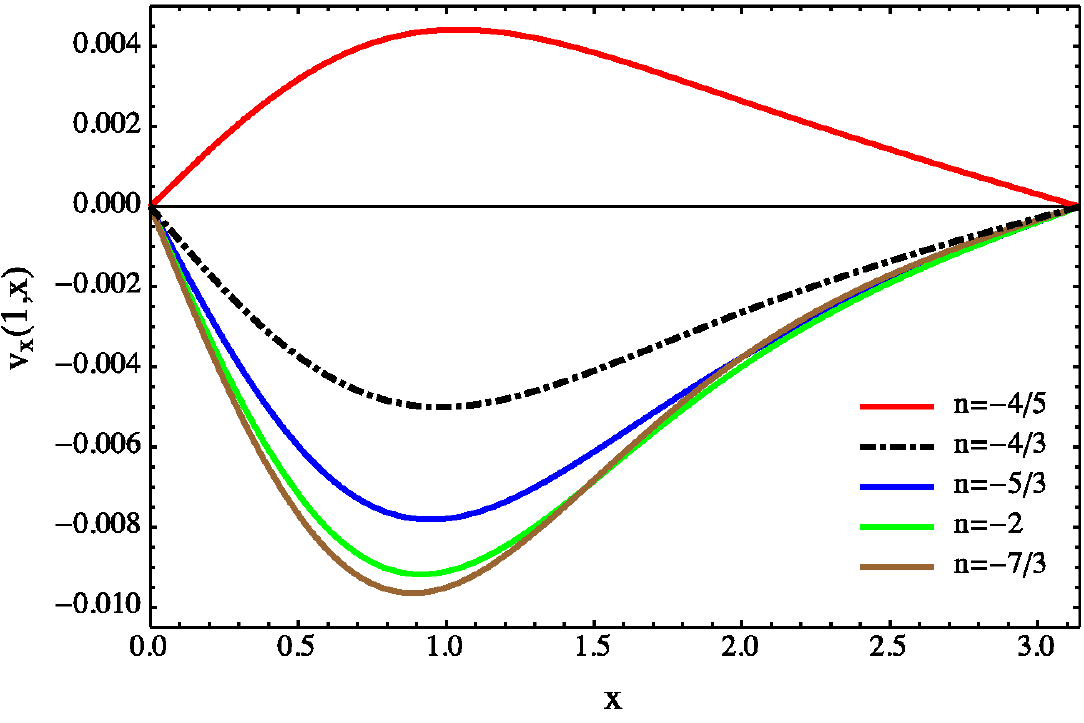}}
			\caption{$v_x$ v.s. $x$ plot at $\tau=1$ with different values of $n$.  The solid curves from bottom to top at $x=1$ correspond to $n=-7/3$(brown), $n=-2$(green), $n=-5/3$(blue), and $n=-4/5$(red), respectively. The dashed curve corresponds to $n=-4/3$(black).}\label{v_x_plot_diffn}
	\end{minipage}
	\hspace {0.5cm}
	\begin{minipage}{8cm}
			{\includegraphics[width=7.5cm,height=5cm,clip]{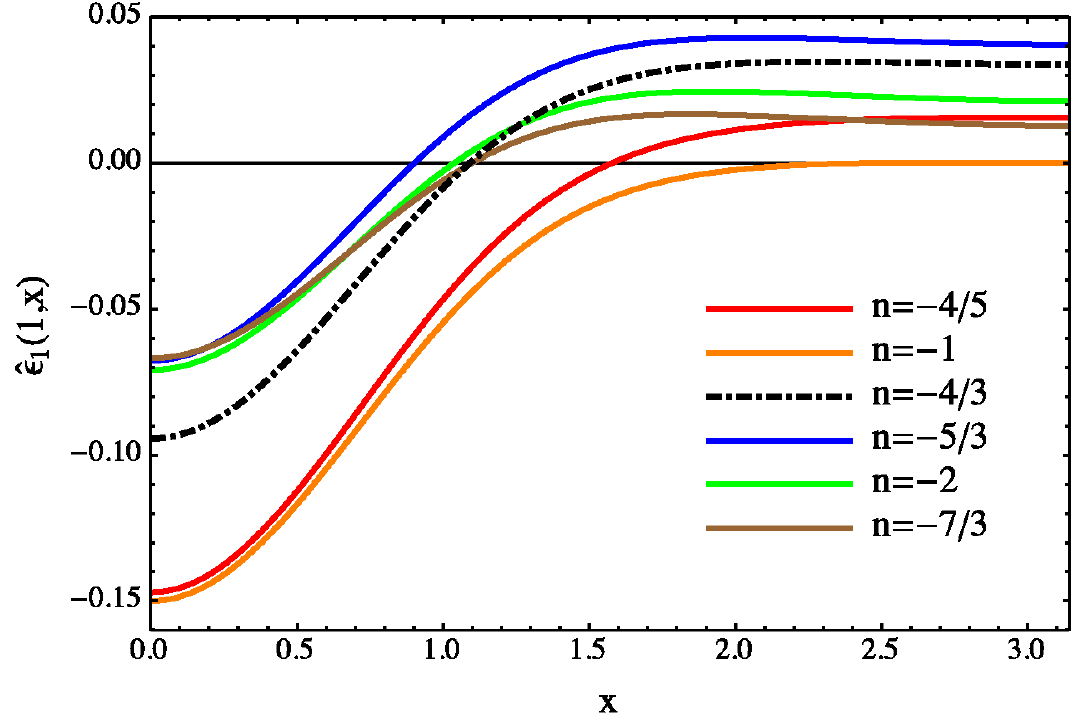}}
			\caption{$\hat{\epsilon}_1=\epsilon_1/\epsilon_c $ v.s. $x$ plot at $\tau=1$ with different values of $n$. The solid curves from top to bottom at $x=0$ correspond to $n=-7/3$(brown), $n=-2$(green), $n=-5/3$(blue), $n=-1$(orange), and $n=-4/5$(red), respectively. The dashed curve corresponds to $n=-4/3$(black).}
			\label{E1_x_plot_diffn}
	\end{minipage}
\end{figure}

In Fig.\ref{v_tau_plot_diffn}, Fig.\ref{E1_tau_plot_diffn}, Fig.\ref{v_x_plot_diffn}, 
and Fig.\ref{E1_x_plot_diffn}, we make further comparisons for $v_x(\tau,x)$ and
 $\epsilon_1(\tau,x)$ at either fixed $x$ or fixed $\tau$ with different values of $n$ 
especially for the cases when $n<-1$. Since $v_x$ and $\epsilon_1$ are even functions 
with respect to $x$, we only plot the results for $x\geq 0$ in Fig.\ref{v_x_plot_diffn} 
and Fig.\ref{E1_x_plot_diffn}. As shown in Fig.\ref{v_tau_plot_diffn}, 
when $|n|$ increases for $n<-1$, $|v_x|$ at fixed $x$ becomes smaller in late
 times due to faster decay of the magnetic field. However, the $|v_x|$ with larger $|n|$ 
may increases in early times. In general, there exists no substantial hierarchy of $|v_x|$ 
at fixed $x$ with different values of $n$ in early times. Furthermore, as shown in 
Fig.\ref{v_x_plot_diffn} at $\tau=1$, we find that the $|v_x|$ first increase from $x=0$ 
and then turn over at intermediate $x$ and gradually decrease with $x$. For $n=-4/5$, 
the velocity profile has a similar shape compared with the cases for $n>-1$, while the
 direction becomes positive. In contrast, according to Fig.\ref{E1_tau_plot_diffn}, the
 increase of $|\epsilon_1|$ at $x=0$ with smaller $|n|$ is observed for almost all times
 except for $\tau\approx 1$ as further shown in Fig.\ref{E1_x_plot_diffn}. In 
Fig.\ref{E1_x_plot_diffn}, we find that the correction on energy density could become
 positive at larger $x$, whereas there exists no clear hierarchy in such a regime.
 Nonetheless, we should mention the caveat that $v_x$ and $\epsilon_1$ shown in 
Fig.\ref{v_x_plot_diffn} and Fig.\ref{E1_x_plot_diffn} are unreliable near $|x|=\pi$,
 where the approximation of the magnetic field in Fourier series breakdowns. In the 
fringes $|x|\approx\pi$, the Gaussian function continues decreasing but the Fourier 
series turnovers as illustrated in Fig.\ref{B_field_plot}. In fact, when plotting the $v_x$ 
and $\epsilon_1$ at fixed $\tau$ from $x=0$ to $x=2\pi$, we find that $v_x$ and $\epsilon_1$ 
becomes odd and even functions with respect to $x=\pi$ since they are formed by sine 
and cosine series, respectively. We may still conjecture the qualitative behaviors near 
the fringes based on our approximated solutions. In Fig.\ref{v_x_plot_diffn}, our approximated
 solutions of $|v_x|$ gradually decreases with $x$ and reach zero at $x=\pi$, while the 
genuine solutions in principle should decay slower and asymptotically coincide with zero
 at $x=\infty$. The qualitative behavior of $\epsilon_1$ near the fringes is more difficult 
to analyze because it depends on the competition between $u_x$ and $B_y$, in which the
 former causes suppression for $n<-1$ and the latter yields enhancement from the second 
equation of (\ref{two_cons_xdep}). Note that $u_x=-v_x$ in our convention and thus
 $\partial_{\tau}u_x$ and $-u_x$ are negative for $n>-1$. We may speculate that the effect
 coming from $u_x$ dominates the one from $B_y$ for $n<-1$ near the fringes and yields 
the suppression of $\epsilon_1$ in large $|x|$. For $n>-1$, the situation is more oblique, where
 both $u_x$ and $B_y$ result in the rise of $\epsilon_1$, which may imply the presence of 
instability near the fringes.  


\begin{figure}[t]
	\begin{minipage}{8cm}
		{\includegraphics[width=7.5cm,height=5cm,clip]{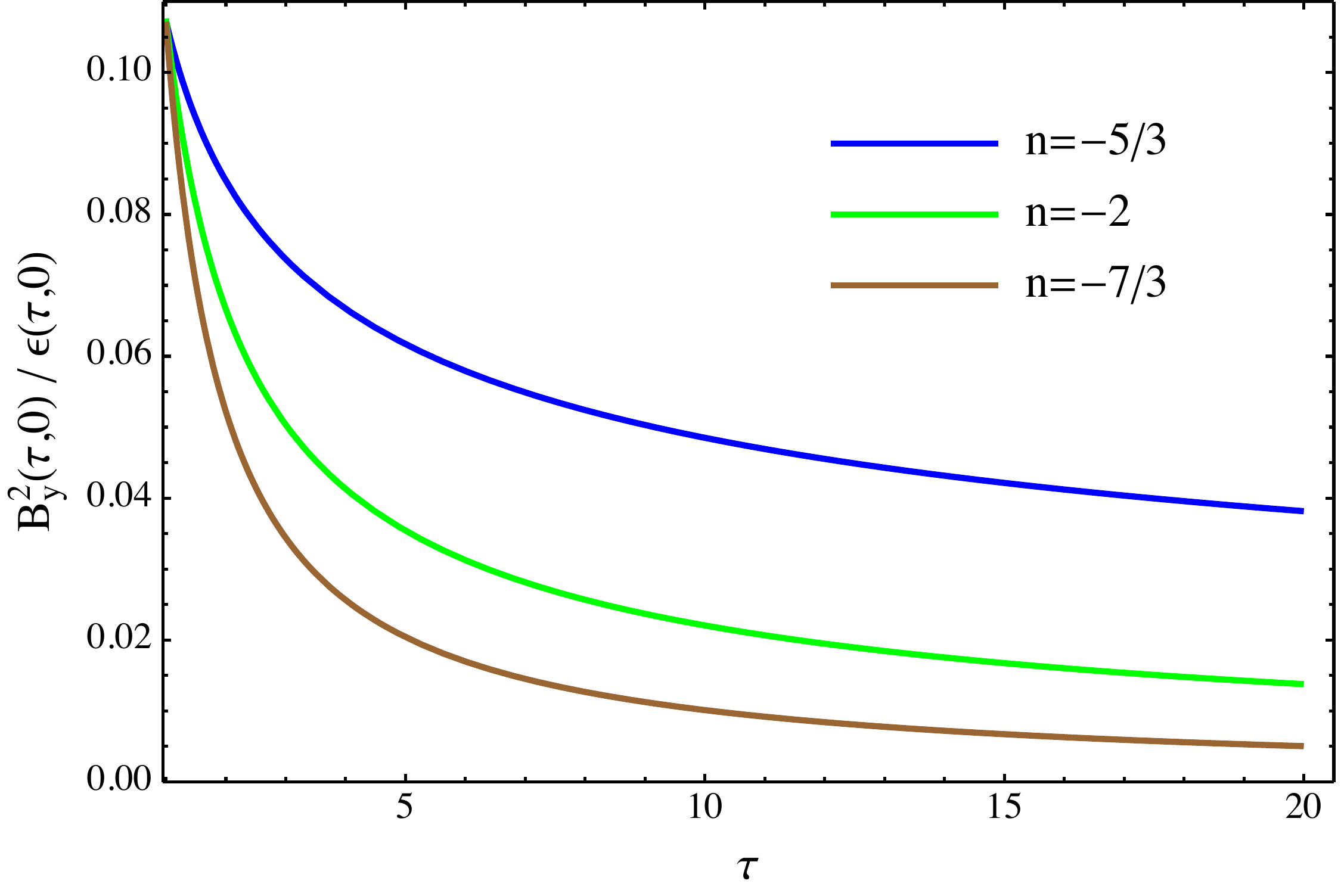}}
		\caption{$B_y^2/\epsilon$ as a function of $\tau$ at $x=0$ with different values of $n$. 
			The solid curves from bottom to top at $x=0$ correspond to $n=-7/3$(brown), $n=-2$(green), 
			and $n=-5/3$(blue), respectively. We have chosen $B_y^2(0,x)/\epsilon_c=0.1$.}
		\label{B_eps_tau}
	\end{minipage}
	\hspace {0.5cm}
	\begin{minipage}{8cm}
		{\includegraphics[width=7.5cm,height=5cm,clip]{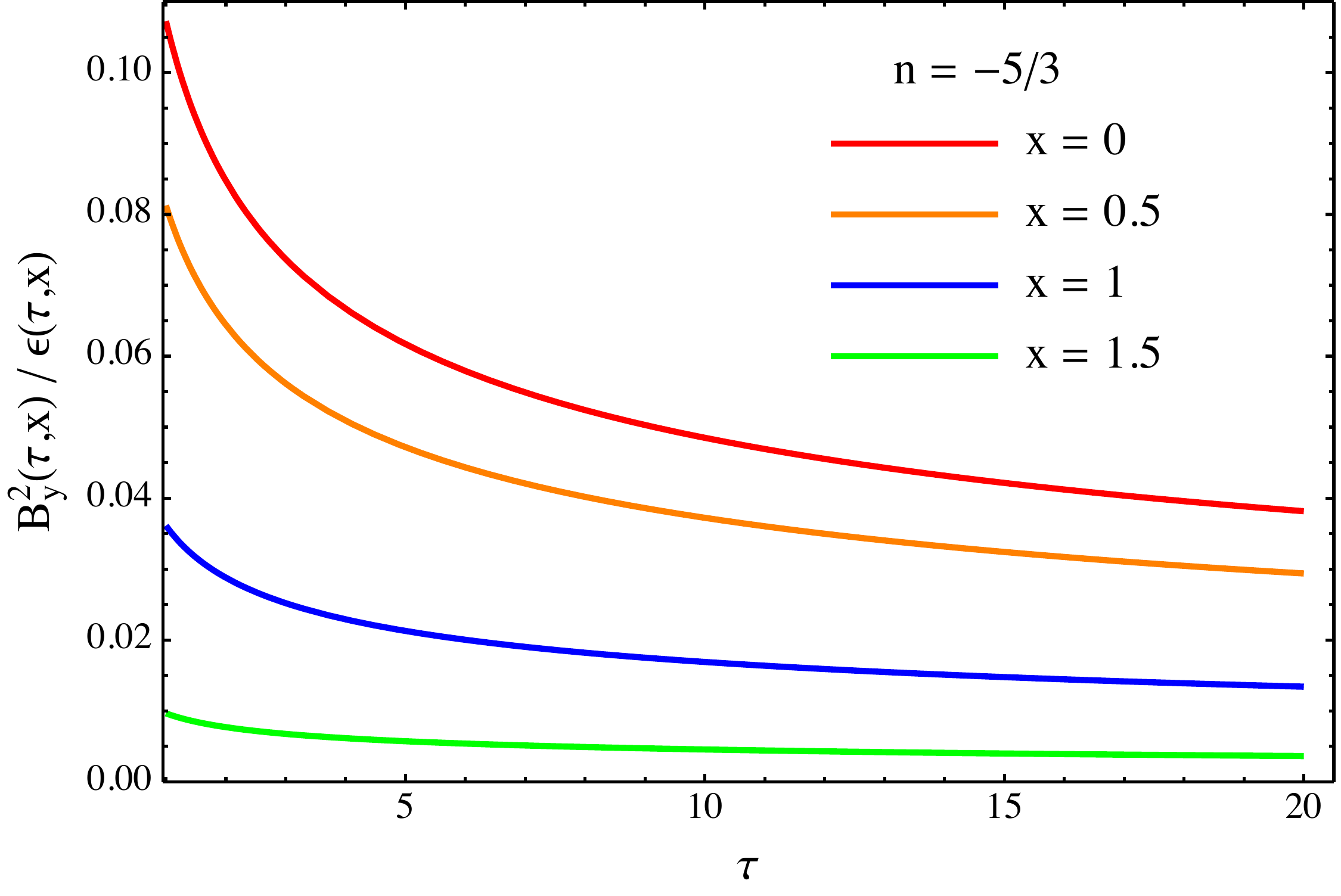}}
		\caption{$B_y^2/\epsilon$ as a function of $\tau$ with different values of $x$. 
			The solid curves from bottom to top correspond to $x=1.5$(green), 
			$x=1$(blue), $x=0.5$(orange), and $x=0$ (red), respectively. We have chosen $B_y^2(0,x)/\epsilon_c=0.1$.}
		\label{B_eps_x}
	\end{minipage}
\end{figure}

Despite the conjectures of the asymptotic behaviors of the solutions near the fringes, we focus on the central region to qualitatively analyze the physics behind the transverse flow generated by magnetic fields. The change of directions of $v_x$ with distinct values of $n$ may be explained by Lenz's law based on the conservation of magnetic flux. To simplify the conditions, we may consider two extreme cases, which correspond to $n\ll -1 $ and $n\gg -1$. For $n\ll -1$, the time scale of the magnetic field is much shorter than the one for the expanding medium, we thus approximate such a condition as a static medium in the presence of a time-decreasing magnetic field $B_y(t)$ with a Gaussian distribution in $x$. The total magnetic flux going through the medium now drops with respect to time. The medium is thus pushed inward to the central region $x=0$ in order to preserve the flux. On the contrary, for $n\gg -1$, the magnetic field decays much slower than the expansion of the medium. We thus approximate the situation with the presence of a static magnetic field $B_y(x)$ as a Gaussian function of $x$ in a medium expanding along the $z$ direction. In such a case, the total magnetic flux of the medium increases with respect to time. To reduce the flux, the medium hence expands along the $\pm x$ directions. The case for $n=-1$ may correspond to the situation in which the magnetic flux is balanced by the expansion of the medium and the decrease of the magnetic field, which thus results in the absence of transverse flow. Note that the medium here can only change the flux via the expansion or compression in the transverse direction due to the absence of induced electromagnetic fields and currents. On the other hand, the correction on the energy density is affected by both the change of fluid velocity and magnetic field, which varies case by case with different values of $n$. Nevertheless, for the peculiar case $n=-1$, one finds that the decrease of the magnetic field always reduces the energy density.


Before we end this section, we also plot the ratio $B_y^2/\epsilon$ as a function of $\tau$ 
for different $n$ and $x$ shown in Fig. \ref{B_eps_tau} and \ref{B_eps_x}, where $\epsilon$ is
the total fluid energy density. 
Since in relativistic heavy ion collisions, the energy density of magnetic fields is expected to decay
much faster than the fluid energy density, in Fig.  \ref{B_eps_tau} we only plot the case $n < -4/3 $.
For a smaller value of $n$, i.e. the magnetic field decays faster, the ratio is smaller. Similarly, in Fig. \ref{B_eps_x} 
the absolute value of $B_y^2$ becomes smaller when $x$ increases, but the fluid energy density is approximately homogeneous. Therefore, the ratio reduces when $x$ increases. 


\section{Conclusions and Outlook}
In this work, we study a toy model of magnetohydrodynamics in the presence of a transverse external 
 magnetic field with spacetime dependence under the Bjorken expansion. In our setup, the medium is boost-invariant along the $z$ direction and the
 magnetic field as a function of $\tau$ and $x$ points along the $y$ direction. We obtain the 
leading-order solutions in the weak-field approximation, where both the energy density and fluid 
velocity are modified. Particularly, the spatial
 gradient of the magnetic field engenders transverse flow parallel or anti-parallel to $x$, 
while the direction and magnitude of flow are determined by the time evolution of the magnetic field.
 For the magnetic field following power-law decay in proper time such that $|{\bf B}|^2\sim \tau^n$, 
the transverse
 flow propagates inward for $n<-1$ and outward for $n>-1$ based on the conservation of magnetic flux 
in the expanding medium. The flow vanishes for $n=-1$, which corresponds to the case such that the 
longitudinal expansion of the medium compensates the decrease of magnetic field. In addition, the energy 
density is generally reduced in the central region, while it can be enhanced in the outskirts depending 
on the competing effects between the transverse velocity and the magnetic field. 

In general, our study in simple setup may provide better understandings for the influence from 
spatial gradient of magnetic fields on magnetohydrodynamics. Although we choose the Gaussian
 distribution as one particular example for the space-dependent magnetic field, the same approach
 can be applied to other spatial distribution given that the interested regime can be approximated by
 Fourier decomposition. Since the analytic expressions of each moment is found, one can directly 
compute the transverse velocity and correction on energy density by just inputting the Fourier 
coefficients. In the end of conclusions, we would like to reemphasize that this study is a theoretical
 discussion, which may be far away from phenomenology in heavy ion collisions. Our solutions might 
be close to the 2+1 dimensional Bjorken flow with
transverse magnetic fields near the initial time when the weak-field approximation becomes valid 
but the transverse flow led by the medium expansion is not fully developed. Here we may further address the validity of our weak-field expansion compared to the practical conditions in heavy ion collisions. According to the numerical simulations \cite{Deng:2012pc,Roy:2015coa}, the magnetic fields generated in peripheral collisions in RHIC at $\tau=0$ are about $|eB_y|\approx 5\sim 10$ $m_{\pi}^2$, whereas the magnitudes may drop to ten times smaller at $\tau\approx 0.6$ fm as the onset of hydrodynamic evolution. Assuming the initial temperature of the QGP is about $T_c=\epsilon_c^{1/4}\approx 300$ MeV, one finds $B_y^2/\epsilon_c\approx 0.17\sim 0.68$ by taking $m_{\pi}\approx 150$ MeV and $e^2=4\pi\alpha=4\pi/137$. As a result, the weak-field expansion in our calculations could be a legitimate approximation for the magnetic fields generated in RHIC. It is helpful to
use our analytical results to test real numerical MHD in the future.           

On the other hand, our study can be generalized along many directions. We may couple the
 conservation equation to electromagnetic currents, which is essential for analyzing the presence
 of chemical potential or chiral anomalous effects. To make further connection with heavy ion collisions,
 the transverse expansion of the medium by itself should be included. The anisotropic flow should
 be simultaneously affected by the expansion of the medium along both the longitudinal and transverse
 directions and also the spacetime-varying magnetic field. Furthermore, 
the viscous effect could be involved as well. How significant the modification from magnetic fields on the flow harmonics measured in experiments is thus relies on the full numerical simulations, which could be affected by the initial conditions chosen for simulations as well. Although the transverse flow led by a inhomogeneous magnetic field shown in this paper is symmetric with respect to the $y$ axis, the flow pattern could become asymmetric from an asymmetric distribution of $B_y$ or the presence of $B_x$. Based on the event-by-event fluctuations \cite{Deng:2012pc}, the magnitude of $B_x$ could be comparable with the magnitude of $B_y$. Consequently, except for the even harmonics, the flow engendered by spatial inhomogeneity of magnetic fields may possibly affect the odd harmonics in heavy ion collisions as well. On the other hand, in order to gain more insights for 
the early-time physics, one may have to seek the next leading-order solutions 
to incorporate nonlinear effects of strong magnetic fields.    

\section*{Acknowledgments}
S.P. is supported by the Alexander von Humboldt
Foundation, Germany and D.Y. is supported by the RIKEN Foreign Postdoctoral Researcher program.

\appendix

\section{rapidity dependence \label{rapidity}}
The magnetic field now is assumed to depend on proper time $\tau$ and rapidity $\eta=\frac{1}{2}\log\left(\frac{t+z}{t-z}\right)$. Up to the leading-order correction from the magnetic field, we introduce the following setup
\begin{eqnarray}
{\bf B}=\lambda B_y(\tau,\eta)\hat{y}, \quad \epsilon=\epsilon_0(\tau)+\lambda^2 \epsilon_1(\tau,\eta),\quad u_{\mu}=(1,0,0,\lambda^2 u_{\eta}(\tau,\eta)),
\end{eqnarray}
where $\lambda$ is an expansion parameter. In the end, after finding the perturbative solution, we may simply set $\lambda=1$. By taking $\epsilon_0=\epsilon_c/\tau^{4/3}$, one finds two conservation equations up to $\mathcal{O}(\lambda^2)$,
\begin{eqnarray}\nonumber\label{two_cons}
&&
\partial_{\tau}\epsilon_1+\frac{4\epsilon_1}{3\tau}-\frac{4\epsilon_c\partial_{\eta}u_{\eta}}{3\tau^{10/3}}+B_y\partial_{\tau}B_y+\frac{B_y^2}{\tau}=0,
\\
&&
\partial_{\eta}\epsilon_1-\frac{4\epsilon_c\partial_{\tau}u_{\eta}}{\tau^{4/3}}
+\frac{4\epsilon_cu_{\eta}}{3\tau^{7/3}}+3B_y\partial_{\eta}B_y=0.
\end{eqnarray}
Combining two coupled differential equations above, one derives a partial differential equation simply depending on $u_{\eta}$,
\begin{eqnarray}\label{equ}
\partial_{\eta}^2 u_{\eta}-u_{\eta}-3\tau^2\partial^2_{\tau}u_{\eta}+\tau\partial_{\tau}u_{\eta}+
\frac{3\tau^{7/3}}{4\epsilon_c}\partial_{\eta}\left(B_y^2+\tau\partial_{\tau}B_y^2\right)=0.
\end{eqnarray}
For $B_y=0$, (\ref{equ}) can be solved by separation of variables, where the solution reads
\begin{eqnarray}
u_{\eta}=\sum_{m}\left[C^m_1\cosh(m\eta)+C^m_2\sinh(m\eta)\right]\tau^{2/3}
\left(\tilde{C}_1^m\tau^{\frac{1}{\sqrt{3}}\sqrt{m^2+1}}+\tilde{C}_2^m
\tau^{-\frac{1}{\sqrt{3}}\sqrt{m^2+1}}\right),
\end{eqnarray}
where $m$ could be either real or imaginary numbers. To find the inhomogeneous solution for $B_y\neq 0$, we may rewrite $B_y^2$ into the Fourier series on the bases of the $\eta$-dependence part of the homogeneous solution for $B_y=0$. Considering $B_y$ generated by two nuclei passing each other in heavy ion collisions, which should be an even function in $\eta$ and most dominant at large rapidity after collisions, we may write
\begin{eqnarray}
B^2_y(\tau,\eta)=\sum_k \tilde{B}_k^2(\tau)\cosh(k\eta), 
\end{eqnarray}
where we choose $k$ as real integers. Note that $B_y$ now is more prominent in large rapidity since the magnetic field generated "close" to one of the moving nucleus is stronger. For simplicity, we may further assume that $\tilde{B}^2_k(\tau)=\tau^n B_k^2$ with $B_k$ being constants. Based on the homogeneous solution, we may assume that the inhomogeneous solution takes the form,
\begin{eqnarray}
u_{\eta}(\tau,\eta)=\sum_m \left[a_m(\tau)\cosh(m\eta)+b_m(\tau)\sinh(m\eta)\right].
\end{eqnarray}
Plugging the ansatz above into (\ref{equ}), one finds $a_m(\tau)=0$ and the inhomogeneous solution reads
\begin{eqnarray}
u_{\eta}(\tau,\eta)=\sum_k\left(-\frac{3B_k^2k(1+n)\tau^{n+7/3}}{4\epsilon_c(k^2-(2+n)(4+3n)}\sinh(k\eta)\right),
\end{eqnarray} 
which then gives rise to 
\begin{eqnarray}
\epsilon_1(\tau,\eta)=\sum_k\left\{\frac{3B_k^2}{2}\frac{[k^2-(2+n)^2]\tau^n\cosh(k\eta)}
{8-k^2+10n+3n^2}\right\}.
\end{eqnarray} 
Here we simply set the integration constants to zero, which should be in fact determined by proper initial conditions in the physical problem.
 
\section{Rescaling for broader inhomogeneity \label{rescaling}}
\begin{figure}[t]
	\begin{minipage}{8cm}
			{\includegraphics[width=7.5cm,height=5cm,clip]{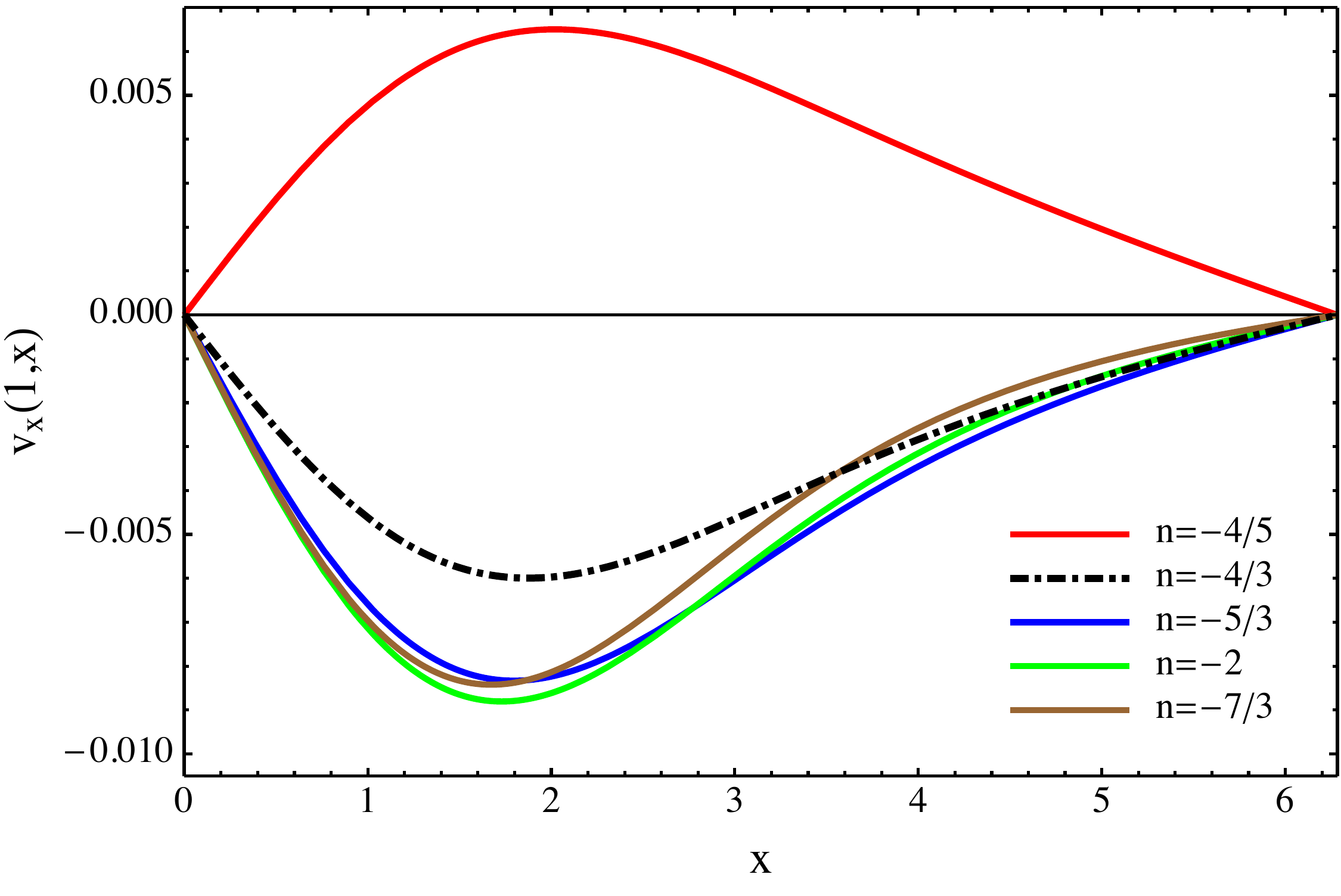}}
			\caption{$v_x$ v.s. $x$ plot for $x_b=2$ at $\tau=1$ with different values of $n$. The color assignments are the same as Fig.\ref{v_x_plot_diffn}.}\label{v_x_plot_diffn_resc}
	\end{minipage}
	\hspace {0.5cm}
	\begin{minipage}{8cm}
			{\includegraphics[width=7.5cm,height=5cm,clip]{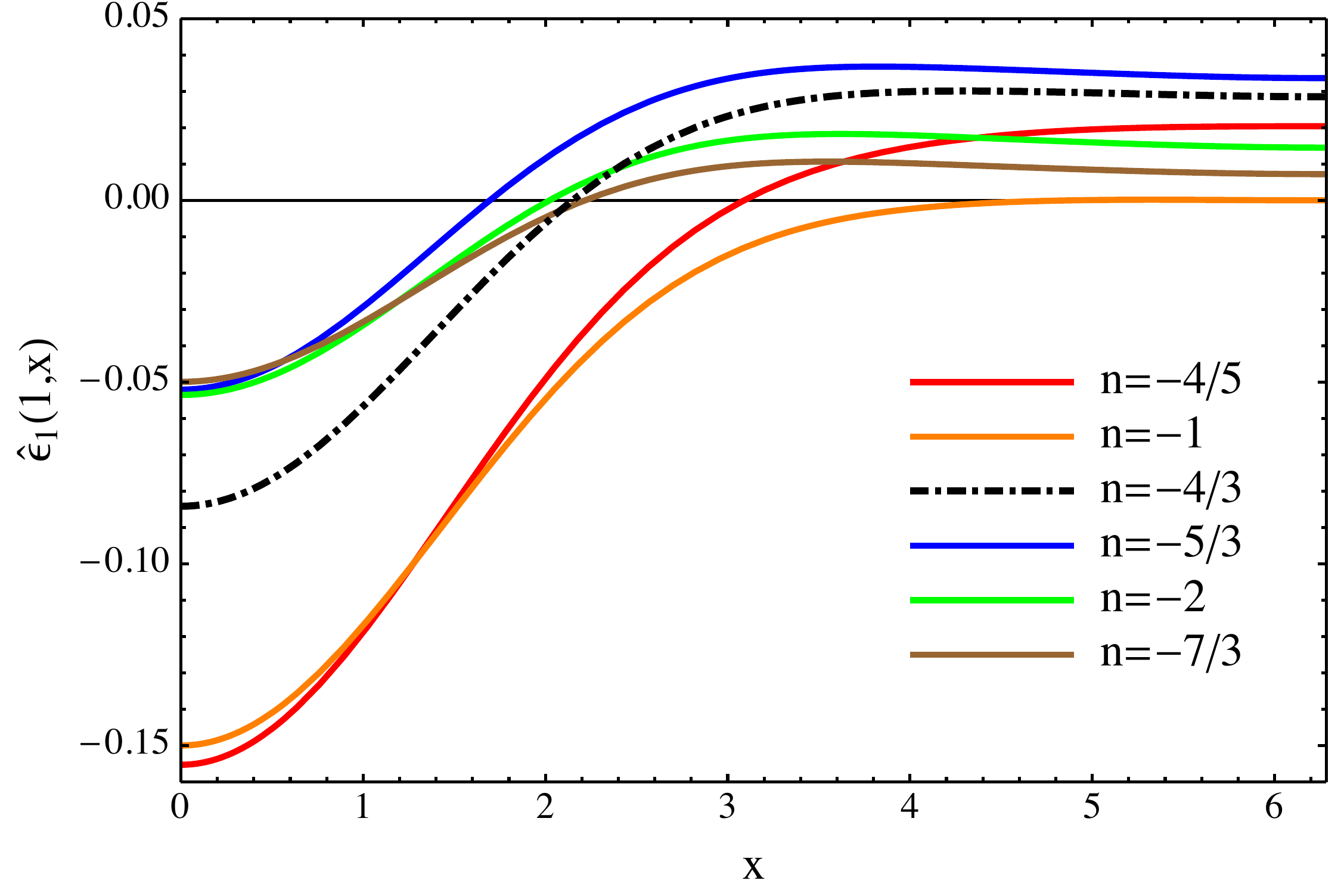}}
			\caption{$\hat{\epsilon}_1=\epsilon_1/\epsilon_c $ v.s. $x$ plot for $x_b=2$ at $\tau=1$ with different values of $n$. The color assignments are the same as Fig.\ref{E1_x_plot_diffn}.}
			\label{E1_x_plot_diffn_resc}
	\end{minipage}
\end{figure}

We consider the magnetic field having large spatial width,
\begin{eqnarray}
B_y^2(\tau,x)=B_c^2\tau^n e^{-x^2/x_b^2},
\end{eqnarray}
where $x_b>1$ is a dimensionless parameter which characterizes the spatial width of the magnetic field as $r_b\sim x_b \tau_0$. Recall that $\tau$ and $x$ are rescaled by $\tau_0$; our original setup is for $x_b=1$. Now, we should work in the rescaled coordinates $(\tau,\bar{x})$ with $\bar{x}=x/x_b$. The magnetic field could be written into the Fourier series,
\begin{eqnarray}
B_y^2(\tau,x)=\tau^n\sum_{\bar{k}}B_{\bar{k}}^2\cos(\bar{k}\bar{x}),
\end{eqnarray}
where $\bar{k}\geq 0$ are integers. The two conservation equations in (\ref{two_cons_xdep}) become
\begin{eqnarray}\nonumber\label{two_cons_rescalex}
&&
\partial_{\tau}\epsilon_1+\frac{4\epsilon_1}{3\tau}-\frac{4\epsilon_c\partial_{\bar{x}}u_{x}}{3x_b\tau^{4/3}}+B_y\partial_{\tau}B_y+\frac{B_y^2}{\tau}=0,
\\
&&
\frac{\partial_{\bar{x}}\epsilon_1}{x_b}-\frac{4\epsilon_c\partial_{\tau}u_{x}}{\tau^{4/3}}
+\frac{4\epsilon_cu_{x}}{3\tau^{7/3}}+\frac{3}{x_b}B_y\partial_{\bar{x}}B_y=0.
\end{eqnarray}
Analogously, combining two equations above yields
\begin{eqnarray}\label{equ_x_rescale}
\frac{\tau^2}{x_b^2}\partial_{\bar{x}}^2 u_{x}-u_{x}-3\tau^2\partial^2_{\tau}u_{x}+\tau\partial_{\tau}u_{x}+
\frac{3\tau^{7/3}}{4x_b\epsilon_c}\partial_{\bar{x}}\left(B_y^2+\tau\partial_{\tau}B_y^2\right)=0.
\end{eqnarray}
Following the same procedure in the computations for $x_b=1$, we find
\begin{eqnarray}
u_x(\tau,\bar{x})=\sum_{\bar{k}}b_{\bar{k}}(\tau)\sin(\bar{k}\bar{x}),
\end{eqnarray}
where $b_{\bar{k}}(\tau)$ is obtained from solving 
\begin{eqnarray}\label{eq_bk_rescale}
(3\tau^2\partial_{\tau}^2-\tau\partial_{\tau}+\frac{\bar{k}^2}{x_b^2}\tau^2+1)b_{\bar{k}}(\tau)+\frac{3B_{\bar{k}}^2}{4x_b\epsilon_c}\bar{k}(n+1)\tau^{n+7/3}=0.
\end{eqnarray} 
Comparing (\ref{eq_bk_rescale}) with (\ref{eq_bk}), the solution can be acquired from the rescaling of the one for $x_b=1$,
\begin{eqnarray}
b_{\bar{k}}(\tau)=b_{k}(\tau)|_{B_k\rightarrow B_{\bar{k}},k->(\bar{k}/x_b)}.
\end{eqnarray}
In fact, $B_{k}=B_{\bar{k}}$ for $k=\bar{k}$, while $\bar{k}/x_b$ may not be integers. Knowing $u_x(\tau,\bar{x})$, one can derive the modification of energy density from  
\begin{eqnarray}\nonumber
\epsilon_1(\tau,\bar{x})&=&\int d\bar{x}\left(\frac{4\epsilon_c}{\tau^{4/3}}x_b\partial_{\tau}u_x-
\frac{4\epsilon_c}{3\tau^{7/3}}x_bu_x-\frac{3}{2}\partial_{\bar{x}}B^2\right)
\\
&=&\epsilon_1(\tau)_{\bar{k}=0}-\sum_{\bar{k}\neq 0}\left(\frac{4\epsilon_c}{\tau^{4/3}\bar{k}}x_b\partial_{\tau}b_{\bar{k}}-
\frac{4\epsilon_c}{3\tau^{7/3}\bar{k}}x_bb_{\bar{k}}+\frac{3}{2}B_{\bar{k}}^2\tau^n\right)\cos(\bar{k}\bar{x}),
\end{eqnarray}  
where $\epsilon_1(\tau)_{\bar{k}=0}$ corresponds to the space-independent part of the solution for $x_b=1$, which remains unchanged after the rescaling. Here we may choose $x_b=2$ as a concrete example. The qualitative behaviors of $v_x$ and $\epsilon_1$ with broader inhomogeneity of $B_y$ are similar to the case for $x_b=1$. In Fig.\ref{v_x_plot_diffn_resc} and \ref{E1_x_plot_diffn_resc}, we plot $v_1$ and $\hat{\epsilon}_1$ at fixed $\tau$ for references.   

\end{document}